\documentclass[11pt,a4paper]{article}
\usepackage{jcappub}
\usepackage{pdflscape}
\usepackage{amsmath}
\usepackage{amssymb}
\usepackage{dcolumn}
\usepackage{bm}
\usepackage{color}
\usepackage{epsfig}
\usepackage{amsfonts}
\usepackage{graphicx}
\usepackage{subfigure}
\usepackage{dcolumn}

\newcommand{\be}{\begin{equation}}
\newcommand{\ee}{\end{equation}}
\newcommand{\bea}{\begin{eqnarray}}
\newcommand{\eea}{\end{eqnarray}}

\def\be{\begin{equation}}
\def\ee{\end{equation}}
\def\bea{\begin{eqnarray}}
\def\eea{\end{eqnarray}}

\begin{document}

	\title{Intermediate accelerated solutions as generic late-time attractors in a modified Jordan-Brans-Dicke theory}

	\author[a]{Antonella Cid}
	
	\author[b]{Genly Leon}
	
	\author[c]{Yoelsy Leyva}

	\affiliation[a]{Grupo de Cosmolog\'ia y Gravitaci\'on GCG-UBB and Departamento de F\'{\i}sica, Universidad del B\'{\i}o-B\'{\i}o, Casilla 5-C, Concepci\'on, Chile}
	\affiliation[b]{Instituto de F\'{\i}sica, Pontificia Universidad  Cat\'olica
		de Valpara\'{\i}so, Casilla 4950, Valpara\'{\i}so, Chile}
	\affiliation[c]{Departamento de F\'{\i}sica, Facultad de Ciencias, Universidad de Tarapac\'a, Casilla 7-D, Arica, Chile}

	\emailAdd{acidm@ubiobio.cl}
	
	\emailAdd{genly.leon@ucv.cl}
	
	\emailAdd{yoelsy.leyva@uta.cl}

	\date{\today}
	\abstract{In this paper we investigate the evolution of a Jordan-Brans-Dicke scalar field, $\Phi$, with a power-law potential in the presence of a second scalar field, $\phi$, with an exponential potential, in both the Jordan and the Einstein frames.  We present the relation of our model with the induced gravity model with power-law potential and the integrability of this kind of models  is discussed when the quintessence field $\phi$ is massless, and has a small velocity. The fact that for some fine-tuned values of the parameters we may
get some integrable cosmological models, makes our choice of potentials very interesting.   We prove that in Jordan-Brans-Dicke theory, the de Sitter solution is not a natural attractor. Instead, we show that the attractor in the Jordan frame corresponds to an ``intermediate accelerated'' solution of the form 	$a(t)\simeq e^{\alpha_1  t^{p_1}}$, as $t\rightarrow \infty$ where $\alpha_1>0$ and  $0<p_1<1$, for a wide range of parameters. 
    Furthermore, when we work in the Einstein frame we get that the attractor is also an ``intermediate accelerated'' solution of the form 	$\mathfrak{a}(\mathfrak{t})\simeq e^{\alpha_2  \mathfrak{t}^{p_2}}$ as $\mathfrak{t}\rightarrow \infty$ where $\alpha_2>0$ and $0<p_2<1$, for the same conditions on the parameter space as in the Jordan frame.
    In the special case of a quadratic potential in the Jordan frame, or for a constant potential in the Einstein's frame,  the above intermediate solutions  are of saddle type. These results were proved using the center manifold theorem, which is not based on linear approximation. Finally, we present a specific elaboration of our extension of the induced gravity model in the Jordan frame, which corresponds to a particular choice of a linear potential of $\Phi$. The dynamical system is then reduced to a two dimensional one, and the late-time attractor is linked with the exact solution found for the induced gravity model. In this example the ``intermediate accelerated'' solution does not exist, and the attractor solution has an asymptotic de Sitter-like evolution law for the scale factor. Apart from some fine-tuned examples such as the linear, and quadratic potential ${U}(\Phi)$ in the Jordan frame, it is true that  ``intermediate accelerated'' solutions are generic late-time attractors in a
modified Jordan-Brans-Dicke theory.}

			\keywords{Modified Gravity, Jordan-Brans-Dicke, Dark Energy, Asymptotic Structure.}
	
	\maketitle



\section{Introduction}

A large amount of research has been devoted to the explanation  of the late-time acceleration of the universe, either by introducing  the concept of Dark Energy or by modifying the gravitational sector itself. Among the simplest
candidates for Dark Energy, one can find canonical scalar fields, phantom fields or the
combination of both fields in a unified quintom model, see \cite{CopelandSamiTsujikawa2006, SetareSaridakis2008}; for the second approach there are several attempts reviewed in \cite{CapozzielloDeLaurentis2011} (see references therein). Despite their interpretation, both approaches can be transformed one into the other, since the crucial issue is just the number of degrees of freedom beyond General Relativity and standard model particles (see
\cite{SahniStarobinsky2006} for a review on such a unified point of view). Finally,
the above scenarios are well-suited not just for late-time implications, likewise for the description of an inflationary stage \cite{NojiriOdintsov2003}.

One example of modified gravitational theory is the so called scalar-tensor theory of gravity \cite{Jordan1959,BransDicke1961, BoisseauEsposito-FaresePolarskiEtAl2000}, in particular the Jordan-Brans-Dicke (JBD) theory \cite{Jordan1959,BransDicke1961}. In this theory  the effective gravitational coupling is time-dependent. The strength of this coupling is determined by a scalar field, the so-called JBD field,  $\Phi \propto G^{-1}$. In modern context, JBD theory appears naturally in supergravity models, Kaluza-Klein theories and in all known effective string actions \cite{Freund1982, AppelquistChodosFreund1987, FradkinTseytlin1985c, FradkinTseytlin1985d, CallanMartinecPerryEtAl1985, CallanKlebanovPerry1986, GreenSchwarzWitten1988, GreenSchwarzWitten1988a}. Furthermore, we can promote the Brans-Dicke (BD) parameter, $\omega_0$, presents in the original theory to a non-constant BD parameter $\omega_0(\Phi),$ and to consider a non-zero
self-interaction potential $U(\Phi),$ even surviving astrophysical tests \cite{BarrowParsons1997, Will1993}.

 In \cite{HrycynaSzydlowski2013} it was investigated the dynamics of the JBD scalar field with a quadratic potential and barotropic matter. The authors used the dynamical systems approach, revealing that the complexity of dynamical evolution, in homogeneous and isotropic cosmological models, depends on the BD parameter, $\omega_0$, and the barotropic matter index,
$w_m$. The authors claim that the quadratic potential function leads naturally to a de Sitter state. The results in \cite{HrycynaSzydlowski2013}  were extended by \cite{HrycynaSzydlowski2013a} for an arbitrary potential function. In \cite{HrycynaSzydlowskiKamionka2014}, it was investigated the observational constraints on the JBD cosmological model using observational data coming from distant supernovae type Ia, the Hubble function, $H(z)$ measurements, information coming from the Alcock-Paczy\'{n}ski test, and baryon acoustic oscillations. 
However, the values found in \cite{HrycynaSzydlowski2013, HrycynaSzydlowski2013a, HrycynaSzydlowskiKamionka2014} for the BD parameter $\omega_0$ are several orders of magnitude lower than the bound $\omega_{0}>4\times 10^4$ imposed by the Solar System tests \cite{Will2014, BertottiIessTortora2003}, and the bounds estimated on the basis of cosmological arguments $\omega_{0}>120$ \cite{AcquavivaBaccigalupiLeachEtAl2005} and $10<\omega_{0}< 10^7$ \cite{NagataChibaSugiyama2004}. 
 This was the main objection to the models  \cite{HrycynaSzydlowski2013, HrycynaSzydlowski2013a, HrycynaSzydlowskiKamionka2014} in \cite{Garcia-SalcedoGonzalezQuiros2015}.  

In this latter paper, the authors stated that the de Sitter solution is an attractor in the Jordan frame of the BD theory  only for the quadratic potential $U(\Phi)\propto \Phi^2$. This result lead them to the claim that de BD cosmology does not have the $\Lambda$CDM model as the universal attractor. Additionally, the authors showed that in the stable de Sitter critical point, as well as in the stiff-matter equilibrium configurations, the dilaton is necessarily massless.  Due to the recent discussions in the literature concerning this topic, we consider it is worthy to investigate the subject further.  In this paper we investigate a JBD scalar field, $\Phi$, with potential ${U}(\Phi)=U_0 \Phi^{2-\frac{\lambda_U}{\gamma}}$, $\gamma^{-1}=\sqrt{\omega_0+\frac{3}{2}}$ (where we have chosen the positive square root by convention), in the presence of a second scalar field, $\phi$, with exponential potential  $V(\phi)=V_0 e^{-\lambda_V \phi}$ to the matter content. Because the addition of $\phi$, we call this scenario a ``modified JBD" theory. 
 We assume that the BD parameter $\omega_0$ is finite, so the limiting case $\omega_0\rightarrow +\infty$ and $\Phi\rightarrow 1$ (we use $8 \pi G=1$), where GR is recovered, is excluded here. 

Several gravity theories consider multiple scalar fields, e.g., assisted inflation scenarios
 	\cite{ChimentoCossariniZuccala1998, GuoPiaoZhang2003, ColeyHoogen2000, CopelandMazumdarNunes1999, HartongPloeghVanRietEtAl2006, MalikWands1999},
 quintom dark energy paradigm \cite{GuoPiaoZhangEtAl2005, ZhangLiPiaoEtAl2006, ArefevaKoshelevVernov2005a, Zhao2006, LazkozLeon2006, Vernov2008, LazkozLeonQuiros2007, SetareSaridakis2008a, CaiSaridakisSetareEtAl2010,  ArefevaBulatovVernov2010, LeonLeyvaSocorro2014}, among
 others \cite{HoogenFilion2000, BagrovObukhov1995, DamourEsposito-Farese1992,  Esposito-Farese1992a, RainerZhuk1996, Wang1998, RodriguesO.SallesShapiroEtAl2011, Tsujikawa2006, GalliKoshelev2011, Clesse2011, XueGaoBrandenberger2012, AlhoMena2014}.  In particular there are some theories where the role of dark matter is played by a scalar field which dynamically behaves as dust during certain epoch in evolution \cite{LimSawickiVikman2010, GaoGongWangEtAl2011, CidLabrana2012}. The main motivation of this work is to analyse if the de Sitter solution represents a natural attractor in the modified JBD model presented.  We investigate the Jordan and Einstein frames, and in both cases we prove that, under the parameter choices $\omega_0>-\frac{3}{2}$ and  $\lambda_U<0$, the late time attractor has an effective equation of state parameter $w_{\text{tot}}=-1$.  This region in the parameter space is compatible with the ranges described by observations \cite{Will2014, BertottiIessTortora2003, AcquavivaBaccigalupiLeachEtAl2005, NagataChibaSugiyama2004, AvilezSkordis2014}.   We prove that in this modified JBD model, the de Sitter solution is not a natural attractor.  Instead, we show that the attractor in the Jordan frame corresponds to an ``intermediate accelerated''  solution of the form 	$a(t)\simeq e^{\alpha_1  t^{p_1}}$ as $t\rightarrow \infty$ with $\alpha_1>0,\; 0<p_1<1$. 
Furthermore, when we work in the Einstein frame we get that the attractor is, as well, an ``intermediate accelerated'' solution of the form 	$\mathfrak{a}(\mathfrak{t})\simeq e^{\alpha_2  \mathfrak{t}^{p_2}}$ as $\mathfrak{t}\rightarrow \infty$ with $\alpha_2>0, \; 0<p_2<1$.

A scale factor of the form $a(t) = \exp\big(At^f\big)$ where $A>0$ and $0<f<1$ was introduced in 
\cite{Barrow1990a, BarrowSaich1990, BarrowLiddle1993} in the context of inflation. Since the expansion of the universe with this scale factor is slower than the de Sitter inflation ($a(t) = \exp
(Ht)$ where $H$ is constant), but faster than the power-law
inflation ($a(t) = t^q$ where $q> 1$), it was called intermediate inflation. 
Intermediate inflationary models arise in the standard inflationary framework as exact cosmological solutions in the slow-roll approximation to
potentials that decay with inverse power-law of the inflaton field \cite{BarrowNunes2007}. These models have been studied in some warm inflationary scenarios  \cite{CampoHerrera2009, CampoHerrera2009a, HerreraVidela2010,  CidCampo2011, CidCampo2012, HerreraOlivaresVidela2013, HerreraOlivaresVidela2013a, Campo2014, HerreraOlivaresVidela2014, HerreraVidelaOlivares2015}.

Intermediate inflation is also found in the context of scalar tensor theories with a variable BD field in the Jordan frame and different matter content. In Ref. \cite{BarrowMimoso1994} a fluid with constant state parameter is considered. In Ref. \cite{Barrow1995} the author takes into account a scalar field as matter source and intermediate inflation is found in  the slow-roll approximation. In the reference \cite{CidCampo2011, CidCampo2012} it was investigated warm intermediate inflation in the JBD theory but formulated in the Einstein frame.  
Since this kind of solutions appear as late-time attractors in our context, we call them ``intermediate accelerated'' solutions. 

We note that under the scalar field rescaling $\sigma=2\sqrt{\omega_0\Phi}$ for $\omega_0>0$ and without the second scalar field $\phi$, we obtain from our model a special case of the so-called induced gravity model, which is integrable for a power-law potential $U(\Phi)$. The general solution of this class of models is known, see for example \cite{KamenshchikPozdeevaTronconiEtAl2014}. After conformal transformation, the induced gravity model with power-law potential becomes a General Relativity model with an exponential potential, that it is integrable as well \cite{KamenshchikPozdeevaTronconiEtAl2014, FreSagnottiSorin2013}.  In this sense, our model can be considered a generalization of the
induced gravity models described above since we have included a new scalar field $\phi$ as the matter source. So it would be interesting to see how the behavior of the solutions for the induced gravity model changes when a small scalar field is added.  In the subsection \ref{Induced_gravity} a discussion of this issue is presented. However, the main purpose of our investigation is not to find analytical solutions but to study the asymptotic behavior of the solutions space of this kind of scenarios without using fine-tuning of the parameters and initial conditions. Dynamical systems theory is a powerful tool for doing this research. Nevertheless, the fact that for some fine-tuned values of the parameters we may get some integrable cosmological models, makes our choice of potentials very interesting.

The paper is organized as follows. In section \ref{SECT:1} the model is presented and the field equations in the Jordan's frame are displayed. In the subsection \ref{Induced_gravity} the relation of our model with the induced gravity model for power-law potential is presented and the integrability of this kind of models is discussed when the quintessence field $\phi$ is massless and has a small velocity, $\dot{\phi}$. In the subsection \ref{SECT:2.1}, the system is written as a dynamical system and the stability of the critical points is discussed. We separated the analysis in two parts: the analysis at the finite region, and the analysis at the infinite region, covering all the possibilities.  In the subsection \ref{JFDiscussion} we present the intermediate accelerated solution as a possible future attractor in the Jordan frame. In subsection \ref{interplay} we investigate a linear potential of $\Phi$, which is equivalent to an extension of the so-called induced gravity model \cite{KamenshchikPozdeevaTronconiEtAl2014, AndrianovCannataKamenshchik2011}. The dynamical system is reduced to a two dimensional one, and the late-time attractor is linked with the solutions found in section \ref{Induced_gravity}. In section \ref{SECT:3} the equations are written in the Einstein's frame, through a conformal transformation. In the subsection \ref{dyn_E} the system is written as a dynamical system and the stability of the critical points is discussed. Special emphasis is given to the possible late time attractors. In subsection \ref{Eframe}  we present the intermediate accelerated solution as a possible future attractor in the Einstein frame. Concluding remarks are given in section \ref{SECT:5}. 
  
\section{Field Equations in the Jordan's frame}\label{SECT:1}

Let us consider the action written in the Jordan frame as given by:
\begin{align}\label{Jordan_action}
S_{JF}&=\int\sqrt{- {g}}\left(\frac{\Phi  {R}}{2}-\frac{\omega_{\text{0}}}{2\Phi} {g}^{\mu\nu} {\partial}_{\mu}\Phi {\partial}_{\nu}\Phi- {U}(\Phi)-\frac{1}{2} {g}^{\mu\nu} {\partial}_{\mu}\phi {\partial}_{\nu}\phi-V(\phi)\right)d^4x,
\end{align}
where $\Phi$ denotes the JBD scalar field, $\omega_0$ is the BD parameter and $\phi$ represents a quintessence scalar field. For the sake of simplicity we restrict our attention to the cases ${U}(\Phi)=U_0 \Phi^{2-\frac{\lambda_U}{\gamma}}$, with $\gamma^{-1}=\sqrt{\omega_0+\frac{3}{2}}$ 
and $V(\phi)=V_0 e^{-\lambda_V \phi}$, but the analysis can be extended to general potentials using similar techniques as in \cite{HrycynaSzydlowski2013, Garcia-SalcedoGonzalezQuiros2015}. $\lambda_U$ and $\lambda_V$ are constants. By construction we have assumed $\gamma$ is positive and finite, it follows $\omega_{0}>-\frac{3}{2}$ (the value $\omega_{0}=-\frac{3}{2}$ gives $\gamma$ infinity). The JBD scalar field $\Phi$ plays the role of an effective Planck mass, consequently we 
assume $\Phi>0$. However, it can asymptotically evolves  to its minimum value $\Phi=0$ as we shall show in the following sections.

By considering a flat Friedmann-Lema\^{\i}tre-Robertson-Walker (FLRW) metric:
\begin{equation}
d  {s}^2=-d {t}^2 +  {a}( {t})^2\left[dr^2 +r^2 \left(d\theta^2 +\sin^2\theta d\varphi^2\right) \right],
\end{equation}
the field equations become 
\begin{subequations}
	\label{EQS_Jordan_1}
	\begin{align}
	&\ddot{\Phi}=\left(\frac{12}{2 \omega_0 +3}-3\right)  {H} \dot{\Phi}+\frac{12  {H}^2 \Phi}{2
		\omega_0 +3}-\frac{2 \Phi  {U}'(\Phi)}{2 \omega_0 +3}-\frac{2 \omega_0  {\dot\Phi}^2}{(2 \omega_0
		+3) \Phi}-\frac{3 {\dot\phi}^2}{2 \omega_0 +3},\label{motion-Phi}\\
	&\ddot\phi=- 3  {H} \dot\phi-V'(\phi), \label{KG_phi_eq}\\
	& 3  {H}^2 \Phi=\frac{\omega_0 {\dot\Phi}^2}{2\Phi}+  {U}(\Phi)+\frac{1}{2}{\dot\phi}^2+V(\phi)-3  {H}\dot\Phi,\label{Fried_JF}\\
	&\dot{ {H}}= \frac{4 \omega_0   {H} \dot\Phi}{(2 \omega_0 +3) \Phi}-\frac{6  {H}^2}{2 \omega_0
		+3}+\frac{ {U}'(\Phi)}{2 \omega_0 +3}-\frac{\omega_0  (2 \omega_0 +1) {\dot\Phi}^2}{2 (2 \omega_0 +3)
		\Phi^2}-\frac{\omega_0  {\dot \phi}^2}{(2 \omega_0 +3) \Phi}, \label{Raych-JF}
	\end{align}
\end{subequations}
where the Hubble expansion rate is given by ${H}=\frac{\dot { {a}}}{ {a}}$ and the dot denotes derivatives with respect to the cosmic time.

By defining the following  effective energy densities $\rho_1, \rho_2$ and the effective pressuress $p_1, p_2$:
\begin{subequations}
	\begin{align}
	&\rho_1=3  {H}^2(1-\Phi) +\frac{\omega_0 {\dot\Phi}^2}{2\Phi}+  {U}(\Phi)-3  {H}\dot\Phi,\\
	&\rho_2=\frac{1}{2}{\dot\phi}^2+V(\phi),\\
	&p_1=H \left(3-\frac{8 \omega_0}{\left(2 \omega_0+3\right) \Phi}\right)
	\dot\Phi+\frac{H^2 \left(\left(6 \omega_0+9\right) \Phi -6 \omega_0+3\right)}{2 \omega_0+3}-\frac{2 U'(\Phi)}{2 \omega_0+3}-U(\Phi)+\nonumber \\ & +\left(\frac{2 \omega_0^2+\omega_0}{\left(2
		\omega_0+3\right) \Phi^2}-\frac{\omega_0}{2 \Phi}\right) \dot{\Phi}^2+\left(\frac{2 \omega_0}{\left(2 \omega_0+3\right) \Phi}-1\right) \dot\phi^2,\\
	& p_2=\frac{1}{2}{\dot\phi}^2-V(\phi),
	\end{align}
\end{subequations}
the system \eqref{EQS_Jordan_1} can be written as
\begin{subequations}
	\label{Jordan_EQS_1}
	\begin{align}
	& \dot {\rho}_1 +3 {H}( {\rho}_1+ {p}_1)=0,\\
	& \dot {\rho}_2 +3 {H}(\rho_2+p_2)=0,\\
	&  {H}^2=\frac{1}{3}\left( {\rho}_1+ {\rho}_2\right),\\
	&\dot{ {H}}=-\frac{1}{2}\left( {\rho}_1+ {p}_1+ {\rho}_2+ {p}_2\right).
	\end{align}
\end{subequations}
The above phenomenological definitions of the energy densities are not unique, specially if an interaction term between both fields is considered  \cite{FaraoniDentSaridakis2014}.

\subsection{Relation with the induced gravity model}\label{Induced_gravity}
							
Let us observe that by setting $\rho_{\phi}:= \frac{1}{2}\dot \phi +V(\phi)=0$, we obtain from \eqref{Jordan_action} the so-called induced gravity model \cite{KamenshchikPozdeevaTronconiEtAl2014, AndrianovCannataKamenshchik2011}:
\begin{align}
S_{IG}=\int\sqrt{- {g}}\left(W(\sigma)  {R}-\frac{1}{2} {g}^{\mu\nu} {\partial}_{\mu}\sigma {\partial}_{\nu}\sigma- {U}(\sigma^2)\right)d^4x,	
\end{align}
under the choices $\sigma=2\sqrt{\omega_0\Phi}$ and $W(\sigma)=\frac{\sigma^2}{8 \omega_0}$, given $\omega_0>0$. 
This model admits exact solutions that we want to discuss in the following. 

Starting with 
${U}(\Phi)=U_0 \Phi^{2-\frac{\lambda_U}{\gamma}}$, $\gamma^{-1}={\sqrt{\omega_0+\frac{3}{2}}}$ (we have chosen the positive square root by convention) and choosing the parameters $\lambda_U=\gamma$ we obtain the potential $$U=\frac{\gamma ^2 U_0 \sigma^2}{4-6 \gamma^2}.$$
In order for $\sigma$ to be real we have chosen  $\omega_0>0$ which implies  $0<\gamma<\sqrt{\frac{2}{3}}$.

Introducing the parametrization \cite{KamenshchikPozdeevaTronconiEtAl2014}
	\begin{subequations}
	\label{eqC2}
	\begin{align}
	a=\sigma^{-1} \exp(u+v),\\
	\sigma=\exp(A(u-v)),
	\end{align}
	\end{subequations}
where $A$ is a constant to be specified, the Friedman equation \eqref{Fried_JF} for $\rho_{\phi}=0$  becomes
\begin{align}\label{Fried_IG_1}
\left(2 A^2-3 \gamma ^2\right)   \dot u^2-2 \left(2 A^2+3 \gamma ^2\right) \dot u \dot v+\left(2 A^2-3 \gamma^2\right) \dot v^2+\gamma ^2
   U_0=0.
\end{align}
Choosing the constant $A=\sqrt{\frac{3}{2}}\gamma$,  \eqref{Fried_IG_1} transforms to (see, e.g., similar equations (28) in \cite{KamenshchikPozdeevaTronconiEtAl2014} and (2.24) in \cite{AndrianovCannataKamenshchik2011}):
\begin{align}\label{light-cone-form}
	\dot u \dot v=\frac{U_0}{12}. 
\end{align}
Substituting  the expressions
\begin{align}
\ddot v =   -\frac{U_0 \ddot u}{12 \dot u^2},\; \dot v= \frac{U_0}{12 \dot u},
\end{align}	
the Raychaudhuri equation \eqref{Raych-JF}  becomes
\begin{align}
3 \gamma ^2 \left(12
   {\dot u}^2+U_0\right)^2+
   4 \sqrt{6} \gamma  \left(12
   {\dot u}^2+U_0\right)
   \left(-3 {\ddot u}-12
   {\dot u}^2+U_0\right)\nonumber \\ +6
   \left(U_0-12
   {\dot u}^2\right) \left(-4
   {\ddot u}-12
   {\dot u}^2+U_0\right)=0,
\end{align}
and the equation of motion for the scalar field \eqref{motion-Phi}, now reduces to	
\begin{align}
-\frac{\gamma ^3 \left(12
   {\dot u}^2+U_0\right)
   e^{\sqrt{6} \gamma 
   (u-v)} \left(-4
   \sqrt{6} {\ddot u}+12
   \left(\gamma
   -\sqrt{6}\right)
   {\dot u}^2+\left(\gamma
   +\sqrt{6}\right)
   U_0\right)}{96
   \left(3 \gamma ^2-2\right)
   {\dot u}^2}=0.
\end{align}	
Since $\gamma$ is nonzero, it follows that both equations are simultaneously satisfied if and only if 
\begin{align}\label{final_u}
\ddot u=\frac{12 \left(\gamma
   -\sqrt{6}\right)
   \dot u^2+\left(\gamma
   +\sqrt{6}\right)
   U_0}{4 \sqrt{6}}.
\end{align}  
The equation \eqref{final_u} admits the general solution :
	\begin{subequations}
			\begin{equation}
			u(t)=\left\{\begin{array}{cc}
			c_2-\frac{2 \ln \left(\cosh \left(\frac{\sqrt{6-\gamma ^2}
   \sqrt{U_0} \left(24 c_1+t\right)}{2
   \sqrt{2}}\right)\right)}{\sqrt{6} \gamma -6}, & \gamma^2<6   \\
	c_2-\frac{2 \ln \left(\cos \left(\frac{\sqrt{\gamma ^2-6}
   \sqrt{U_0} \left(24 c_1+t\right)}{2
   \sqrt{2}}\right)\right)}{\sqrt{6} \gamma -6}, & \gamma^2>6  .
	\end{array}
	\right.
				\end{equation}
		Substituting the result for $u$ on the equation \eqref{light-cone-form}, and integrating out the resulting equation
		we obtain 
				\begin{equation}
				v(t)= \left\{ \begin{array}{cc} c_3+ \frac{2 \ln \left(\sinh \left(\frac{\sqrt{6-\gamma ^2} \sqrt{U_0}
   \left(24 c_1+t\right)}{2 \sqrt{2}}\right)\right)}{\sqrt{6} \gamma
   +6}, & \gamma^2<6 \\  c_3+ \frac{2 \ln \left(\sin \left(\frac{\sqrt{\gamma ^2-6} \sqrt{U_0}
   \left(24 c_1+t\right)}{2 \sqrt{2}}\right)\right)}{\sqrt{6} \gamma
   +6}, & \gamma^2>6
	\end{array}
	\right. 
				\end{equation}
	\end{subequations}
	respectively, where $c_1, c_2$ and $c_3$ are integration constants.

Since we have chosen $0<\gamma<\sqrt{\frac{2}{3}}$, given that $\omega_0>0$, hereafter we use the branch given by hyperbolic functions:
\begin{subequations}
\label{eqC9}
\begin{align}
u(t)=c_2-\frac{2 \ln \left(\cosh \left(\frac{\sqrt{6-\gamma ^2}
   \sqrt{U_0} \left(24 c_1+t\right)}{2
   \sqrt{2}}\right)\right)}{\sqrt{6} \gamma -6},\\
v(t)=c_3+ \frac{2 \ln \left(\sinh \left(\frac{\sqrt{6-\gamma ^2} \sqrt{U_0}
   \left(24 c_1+t\right)}{2 \sqrt{2}}\right)\right)}{\sqrt{6} \gamma
   +6}.	
\end{align}
\end{subequations} 
		
Substituting \eqref{eqC9} in \eqref{eqC2} we obtain the solutions:
	\begin{subequations}
	\label{solutions-induced-gravity}
	\begin{align}
	&\sigma(t)=e^{\sqrt{\frac{3}{2}} \gamma 
   \left(c_2-c_3\right)} \sinh
   ^{-\frac{\sqrt{6}\gamma}{\sqrt{6} \gamma
   +6}}(\Delta(t) ) \cosh
   ^{-\frac{\sqrt{6}\gamma}{\sqrt{6} \gamma
   -6}}(\Delta(t) ),\\
	&a(t)=e^{\sqrt{\frac{3}{2}} \gamma  \left(c_3-c_2\right)+c_2+c_3} \sinh ^{\frac{\sqrt{6} \gamma
   +2}{\sqrt{6} \gamma +6}}(\Delta (\tau
   )) \cosh ^{\frac{\sqrt{6} \gamma
   -2}{\sqrt{6} \gamma -6}}(\Delta (\tau
   ))
	,\\
	&H(t)=\frac{\sqrt{U_0
	} \text{csch}(2 \Delta (t)) \left(2 \sqrt{6} \gamma -3 \left(\gamma
   ^2-2\right) \cosh (2 \Delta (t))\right)}{3 \sqrt{2} \sqrt{6-\gamma ^2}},
	\end{align}
	\end{subequations}
	where 
	\begin{equation}
	\Delta(t)=\frac{\sqrt{6-\gamma ^2} \sqrt{U_0} \left(24 c_1+t\right)}{2 \sqrt{2}}. 
	\end{equation}

	\subsubsection{Including a massless scalar field}\label{massless_s_f}
	
The equation of motion \eqref{KG_phi_eq} for a massless scalar field is given by 
\begin{equation}
\ddot\phi +3\frac{\dot a}{a}\dot\phi=0,
\end{equation}
and it admits the solution $\dot\phi=\mu a^{-3}$, where $\mu$ is an integration constant. 	Combining the parametrization 	\eqref{eqC2} with $A=\sqrt{\frac{3}{2}}\gamma,$ 
the Raychaudhuri \eqref{Raych-JF} and the equation of motion for the scalar field \eqref{motion-Phi}  we obtain 
\begin{subequations}
\begin{align}
& \ddot u=\frac{\left(\sqrt{6} \gamma
   +2\right) \left(3 \gamma
   ^2-2\right) \mu^2 \exp
   \left(2 \sqrt{6} \gamma 
   (u -v )-6 u -6
   v \right)}{4 \gamma
   ^2}\nonumber \\ & \qquad  +\left(\sqrt{\frac{3}{2}
   } \gamma -3\right)
   \dot u^2+\left(3
   \sqrt{\frac{3}{2}} \gamma
   +3\right) \dot u
   \dot v-\frac{\gamma 
   U_0}{2
   \sqrt{6}},\\
	& \ddot v =-\frac{\left(\sqrt{6}
   \gamma -2\right) \left(3
   \gamma ^2-2\right) \mu^2
   \exp \left(2 \sqrt{6}
   \gamma  (u -v )-6
   u -6 v \right)}{4
   \gamma ^2}\nonumber \\ & \qquad +\left(3-3
   \sqrt{\frac{3}{2}} \gamma
   \right) \dot u
   \dot v+\left(-\sqrt{\frac{3}
   {2}} \gamma -3\right)
   \dot v^2+\frac{\gamma 
   U_0}{2
   \sqrt{6}},	
\end{align}
\end{subequations}
and the Friedmann equation \eqref{Fried_JF}, assuming that $V(\phi)=0$, i.e., the massless case, now becomes
\begin{subequations}
\begin{align}
&\dot u \dot v= G(u,v),\\
&G(u,v)=\frac{1}{12} \left(\frac{\left(2-3
   \gamma ^2\right) \mu^2 \exp
   \left(2 \sqrt{6} \gamma 
   (u-v)-6 u-6
   v\right)}{\gamma
   ^2}+U_0\right).	
\end{align}
\end{subequations}
Combining the above equations we obtain 
\begin{align}\label{eqC15}
&\frac{3 \left(\sqrt{6} \gamma
   +2\right) \left(3 \gamma
   ^2-2\right) \mu^2 \exp
   \left(2 \sqrt{6} \gamma 
   (u-v)-6 u-6
   v\right)}{\gamma }\nonumber \\ & \qquad -24
   \gamma  \ddot u+12 \gamma 
   \left(\sqrt{6} \gamma
   -6\right) \dot u^2+\gamma 
   \left(\sqrt{6} \gamma
   +6\right) U_0=0,
\end{align}
 which reduces to \eqref{final_u} for $\mu=0$.

Now we want to simplify further the  equations, choosing a new time parameter $\tau$ such that
\begin{equation}
u' v' \dot\tau^2=G(u,v),
\end{equation}
where the comma denotes derivative with respect the new time $\tau$. Thus,
choosing $\dot\tau=\sqrt{G(u,v)}$ we obtain 
\begin{equation}
\label{eqC17}
u'v'=1. 
\end{equation}
The second derivatives with respect to $t$ are given by 
\begin{subequations}
\begin{align}
&\ddot u={\dot\tau}^2 u''+\ddot\tau  u',\\
&\ddot v={\dot\tau}^2 v''+\ddot\tau  v',
\end{align}
\end{subequations}
where 
\begin{subequations}
\begin{align}
&\dot\tau =
   \frac{\sqrt{\frac{\left(2-3
   \gamma ^2\right) \mu^2
   e^{2 \sqrt{6} \gamma  (u-v)-6 (u+v)}}{\gamma ^2}+U_0}}{2
   \sqrt{3}},\\
&	\ddot \tau =
   -\frac{\left(3 \gamma
   ^2-2\right) \mu^2 
   \left(\left(\sqrt{6} \gamma
   -3\right) u'-\left(\sqrt{6} \gamma
   +3\right) v'\right)e^{2 \sqrt{6} \gamma  (u-v)-6 (u+v)}}{12 \gamma
   ^2}.
\end{align}
\end{subequations}
	
Finally, the equation \eqref{eqC15} transforms to 
{\small{
\begin{align}
\label{EqC20}
& u'' \left(2 \left(3
   \gamma ^2-2\right) \mu^2
   e^{2 \sqrt{6} \gamma 
   u}-2 \gamma ^2
   U_0 e^{6 u+2 \left(\sqrt{6}
   \gamma +3\right) v}\right)\nonumber\\ & +\gamma 
   \left(\sqrt{6} \left(3
   \gamma ^2-2\right) \mu^2
   e^{2 \sqrt{6} \gamma 
   u} \left({u'}^2+1\right)+\gamma 
   U_0 \left(\sqrt{6}
   \gamma +\left(\sqrt{6}
   \gamma -6\right) {u'}^2+6\right) e^{6 u+2 \left(\sqrt{6}
   \gamma +3\right) v}\right)=0.
\end{align}
}}	

As in the previous case, since we are interested in the range of parameters $0<\gamma<\sqrt{\frac{2}{3}}$, we omit the solutions of \eqref{EqC20} given in terms of trigonometric functions and we use hyperbolic ones instead. Thus, for the unmodified case $\mu=0$ we recover the exact solution \eqref{eqC9}
\begin{subequations}
\begin{align}
  & u(\tau)=   c_2-\frac{2 \ln
   \left(\cosh
   \left(\Delta(\tau)\right)\right)}{\sqrt{6
   } \gamma -6},\\
	& v(\tau)=c_3+
   \frac{2 \ln \left(\sinh
   \left(\Delta(\tau)\right)\right)}{\sqrt{6
   } \gamma
   +6},
\end{align}
\end{subequations}
where $\Delta(\tau)=\frac{\sqrt{6}}{2}
   \sqrt{6-\gamma^2} \left(2
   c_1+\tau \right)$, defined for $\gamma^2<6$.

We can use the solution \eqref{seed_solution} for constructing an approximated solution for the system when $\mu\neq 0$ is a small parameter, i.e., assuming that the scalar field $\phi$ is massless and has a small velocity $\dot\phi$. 
 
Assuming $\mu \neq 0$, we define
\begin{subequations}
\label{eqTay}
 \begin{align}
 & u(\tau)=U(\tau)+ \mu d_2(\tau)+O(\mu^2),\\
	& v(\tau)=V(\tau) + \mu d_3(\tau) +O(\mu^2),
\end{align}
\end{subequations}
where $U, V$ are the seed solutions when $\mu=0$ given by 
\begin{subequations}
\label{seed_solution}
\begin{align}
  & U(\tau)=   c_2-\frac{2 \ln
   \left(\cosh
   \left(\Delta(\tau)\right)\right)}{\sqrt{6
   } \gamma -6},\\
	& V(\tau)=c_3+
   \frac{2 \ln \left(\sinh
   \left(\Delta(\tau)\right)\right)}{\sqrt{6
   } \gamma
   +6},
\end{align}
\end{subequations}
and $d_2, d_3$ are functions to be specified. Substituting in \eqref{eqTay} and in \eqref{EqC20}, expanding in Taylor's series with respect to the parameter $\mu$ near $\mu=0$, we obtain respectively:
\begin{align}
\mathcal{E}_{11} +\mu \mathcal{E}_{12} +O(\mu^2)=0,\\
\mathcal{E}_{21} +\mu \mathcal{E}_{22} +O(\mu^2)=0,
\end{align}
where the equations $\mathcal{E}_{11}=0,\mathcal{E}_{12}=0,\mathcal{E}_{21}=0,\mathcal{E}_{22}=0$ must be satisfied. 
\begin{subequations}
\begin{align}
&\mathcal{E}_{11}=0 \implies U'(\tau ) V'(\tau )-1=0, \\
&\mathcal{E}_{12}=0 \implies U'(\tau ) {d_3}'(\tau )+V'(\tau ) {d_2}'(\tau )=0,\\
&\mathcal{E}_{21}=0\implies \sqrt{6} \gamma -2 U''(\tau )+\left(\sqrt{6} \gamma -6\right) U'(\tau
   )^2+6=0,\\
&\mathcal{E}_{22}=0 \implies -2 {d_2}''(\tau )+2 \left(\sqrt{6} \gamma -6\right) U'(\tau ) {d_2}'(\tau
   ) \nonumber \\ & +\left(2 \left(\sqrt{6} \gamma +3\right) {d_3}(\tau )+6 {d_2}(\tau )\right)
   \left(\sqrt{6} \gamma -2 U''(\tau )+\left(\sqrt{6} \gamma -6\right) U'(\tau
   )^2+6\right)=0.	
\end{align}
\end{subequations}	
Substituting \eqref{seed_solution} in the above equations makes the equations $\mathcal{E}_{11}=0$ and $\mathcal{E}_{21}=0$ trivially satisfied, and  
the equations $\mathcal{E}_{12}=0$ and $\mathcal{E}_{22}=0$ simplify now to
\begin{subequations}
\begin{align}
&{d_2}''(\tau )= -\sqrt{6}
   \sqrt{6-\gamma ^2} {d_2}'(\tau ) \tanh
   \left(\Delta(\tau)\right),\\
& {d_3}'(\tau )=
  -\frac{\left(\sqrt{6}-\gamma
   \right) {d_2}'(\tau ) \coth
   ^2\left(\Delta(\tau)\right)}{\gamma
   +\sqrt{6}},
\end{align}
\end{subequations} where $\Delta(\tau)=\frac{\sqrt{6}}{2}
   \sqrt{6-\gamma^2} \left(2
   c_1+\tau \right)$, 
with solutions
\begin{subequations}
\label{eqC28}
\begin{align}
&d_2(\tau )= \frac{\sqrt{\frac{2}{3}} f_1 \tanh
   \left(\Delta(\tau)\right)}{\sqrt{6-\gamma
   ^2}}+f_2,\\
&	d_3(\tau )=  \frac{\sqrt{\frac{2}{3}}
   \sqrt{\sqrt{6}-\gamma } f_1 \coth
   \left(\Delta(\tau)\right)}{\left(\gamma
   +\sqrt{6}\right)^{3/2}}+f_3,
\end{align}
where $f_1, f_2, f_3$ are integration constants. 
\end{subequations}
Henceforth, we obtain the first order (in the parameter $\mu$) solution
\begin{subequations}
\label{seed_solution_first_order}
\begin{align}
  & u(\tau)=    c_2-\frac{2 \ln
   \left(\cosh
   \left(\Delta(\tau)\right)\right)}{\sqrt{6
   } \gamma -6} +\mu \left[\frac{\sqrt{\frac{2}{3}} f_1 \tanh
   \left(\Delta(\tau)\right)}{\sqrt{6-\gamma
   ^2}}+f_2\right]+O(\mu^2),\\
	& v(\tau)=c_3+
   \frac{2 \ln \left(\sinh
   \left(\Delta(\tau)\right)\right)}{\sqrt{6
   } \gamma
   +6}+\mu \left[\frac{\sqrt{\frac{2}{3}}
   \sqrt{\sqrt{6}-\gamma } f_1 \coth
   \left(\Delta(\tau)\right)}{\left(\gamma
   +\sqrt{6}\right)^{3/2}}+f_3\right]+O(\mu^2).
\end{align}
\end{subequations}

The relative errors in the approximation of \eqref{seed_solution_first_order} by \eqref{seed_solution} are:
\begin{subequations}
\label{relative-errors}
\begin{align}
& E_r(u(\tau)):=\frac{u(\tau)-U(\tau)}{u(\tau)}=\frac{\mu 
   \left(\frac{\sqrt{\frac{2}{3}} f_1
   \tanh \left(\Delta(\tau)\right)}{\sqrt{6-\gamma
   ^2}}+f_2\right)}{c_2-\frac{2 \ln
   \left(\cosh \left(\Delta(\tau)\right)\right)}{\sqrt{6}
   \gamma -6}}+O\left(\mu ^2\right),\\
& E_r(v(\tau)):=\frac{v(\tau)-V(\tau)}{v(\tau)}=\frac{\mu 
   \left(\frac{\sqrt{\frac{2}{3}}
   \sqrt{\sqrt{6}-\gamma } f_1 \coth
   \left(\Delta(\tau)\right)}{\left(\gamma
   +\sqrt{6}\right)^{3/2}}+f_3\right)}{
   \frac{2 \ln \left(\sinh
   \left(\Delta(\tau)\right)\right)}{\sqrt{6}
   \gamma +6}+c_3}+O\left(\mu ^2\right).
\end{align}
\end{subequations}
Taking the limit $\tau\rightarrow +\infty$ it follows that the above relative errors tend to zero. Thus, the linear terms in $\mu$ in the equation \eqref{seed_solution_first_order} can be made a small percent of the contribution of the zeroth-solutions \eqref{seed_solution} taking $\tau$ large enough. Henceforth, this shows that the behavior of the solutions for the induced gravity model does not change abruptly  when a small massless scalar field, $\phi$, is added to the setup. 

Finally, going back to the original variables, we obtain the solutions: 
\begin{subequations}
\label{cosmol_sol}
\begin{align}
& \sigma(\tau)=e^{\sqrt{\frac{3}{2}} \gamma 
   \left(c_2-c_3\right)} \sinh
   ^{-\frac{\sqrt{6} \gamma }{\sqrt{6}
   \gamma +6}}(\Delta (\tau )) \cosh
   ^{-\frac{\sqrt{6} \gamma }{\sqrt{6}
   \gamma -6}}(\Delta (\tau )) \times \nonumber \\ &  \left[1+\frac{1}{2} \gamma  \mu 
   \left(\sqrt{6} (f_2-f_3)-\frac{4
   f_1 \text{csch}(2 \Delta (\tau ))
   \left(\sqrt{6}-\gamma  \cosh (2
   \Delta (\tau
   ))\right)}{\sqrt{\sqrt{6}-\gamma }
   \left(\gamma
   +\sqrt{6}\right)^{3/2}}\right)\right] +O(\mu^2),
\\	
& a(\tau)= \frac{\sinh ^{\frac{\sqrt{6}
   \gamma +2}{\sqrt{6} \gamma
   +6}}(\Delta (\tau )) \cosh
   ^{\frac{\sqrt{6} \gamma -2}{\sqrt{6}
   \gamma -6}}(\Delta (\tau ))}{K_1^{1/3}} \nonumber \\
&	+\frac{f_1 \mu   \left(2 \sqrt{6}
   \gamma -3 \left(\gamma ^2-2\right)
   \cosh (2 \Delta (\tau ))\right) \sinh
   ^{-\frac{4}{\sqrt{6} \gamma
   +6}}(\Delta (\tau )) \cosh
   ^{\frac{4}{\sqrt{6} \gamma
   -6}}(\Delta (\tau ))}{3
   \sqrt{\sqrt{6}-\gamma } \left(\gamma
   +\sqrt{6}\right)^{3/2}
   K_1^{1/3}}\nonumber \\
& -\frac{f_2 \mu \left(\gamma 
   \left(\gamma  \left(\sqrt{6} \gamma
   +10\right)+2
   \sqrt{6}\right)-12\right) 
   \sinh (2 \Delta (\tau )) \sinh
   ^{-\frac{4}{\sqrt{6} \gamma
   +6}}(\Delta (\tau )) \cosh
   ^{\frac{4}{\sqrt{6} \gamma
   -6}}(\Delta (\tau ))}{4 \left(\gamma
   +\sqrt{6}\right)^2
   K_1^{1/3}} \nonumber \\
&+\frac{f_3 \mu\left(\gamma 
   \left(\gamma  \left(\sqrt{6} \gamma
   +14\right)+10
   \sqrt{6}\right)+12\right)  
   \sinh (2 \Delta (\tau )) \sinh
   ^{-\frac{4}{\sqrt{6} \gamma
   +6}}(\Delta (\tau )) \cosh
   ^{\frac{4}{\sqrt{6} \gamma
   -6}}(\Delta (\tau ))}{4 \left(\gamma
   +\sqrt{6}\right)^2
   K_1^{1/3}} +O(\mu^2),
\\
	&H(\tau)=\frac{\dot\tau a'(\tau)}{a(\tau)}=\frac{\sqrt{G(u(\tau), v(\tau))}\; a'(\tau)}{a(\tau)}\nonumber \\
	&=\frac{\sqrt{F(\tau )}
   \text{csch}(2 \Delta (\tau ))
   \left(4 \sqrt{3} \gamma -3
   \sqrt{2} \left(\gamma
   ^2-2\right) \cosh (2 \Delta
   (\tau ))\right)}{6
   \sqrt{6-\gamma ^2}} \nonumber \\ 
	&+ \frac{2 \gamma  f_1 \mu  \coth (2
   \Delta (\tau )) \text{csch}(2
   \Delta (\tau )) \left(\left(30-9
   \gamma ^2\right) F(\tau
   )+\left(7 \gamma ^2-18\right)
   U_0\right)}{\sqrt{3}
   \left(\gamma +\sqrt{6}\right)
   \left(\gamma ^2-6\right)
   \sqrt{F(\tau )}} \nonumber \\ 
	&+\frac{\sqrt{2}
   f_1 \mu  \text{csch}^2(2 \Delta
   (\tau )) \left(\left(2 \gamma
   ^4-9 \gamma ^2+18\right) F(\tau
   )+\left(-\gamma ^4+\gamma
   ^2-6\right)
   U_0\right)}{\left(\gamma
   +\sqrt{6}\right) \left(\gamma
   ^2-6\right) \sqrt{F(\tau
   )}} \nonumber \\
	&-\frac{\sqrt{2} \left(\gamma
   ^4-5 \gamma ^2+6\right) f_1 \mu
    \cosh (4 \Delta (\tau ))
   \text{csch}^2(2 \Delta (\tau ))
   (U_0-F(\tau
   ))}{\left(\gamma
   +\sqrt{6}\right) \left(\gamma
   ^2-6\right) \sqrt{F(\tau
   )}} \nonumber \\
		&+\frac{\alpha_2 \mu  \text{csch}(2 \Delta
   (\tau )) (F(\tau )-U_0)
   \left(\sqrt{6} \left(\gamma
   ^2-2\right) \cosh (2 \Delta
   (\tau ))-4 \gamma \right)}{2
   \gamma  \sqrt{F(\tau
   )}} \nonumber\\
	&+\frac{\alpha_3 \mu 
   (F(\tau )-U_0)
   \left(\sqrt{6} \left(\gamma
   ^2-2\right) \coth (2 \Delta
   (\tau ))-4 \gamma  \text{csch}(2
   \Delta (\tau ))\right)}{4 \gamma
    \sqrt{F(\tau )}},
\\
	\phi & =\phi_0+\int_{\tau_0}^\tau \frac{\mu}{\sqrt{G(u(\tau'), v(\tau'))}a(\tau')^3 } d\tau'\nonumber\\
	& = \phi_0 + \int_{\tau_0}^\tau \left(\frac{2 \sqrt{3} \mu  K_1 \sinh
   ^{\frac{12}{\sqrt{6} \gamma
   +6}-3}(\Delta (\tau' )) \cosh
   ^{-\frac{12}{\sqrt{6} \gamma
   -6}-3}(\Delta (\tau'
   ))}{\sqrt{F(\tau')}}\right) d\tau' +O(\mu^2),
	\\
	&t=t_0 + \int_{\tau_0}^\tau \frac{d\tau'}{\sqrt{G(u(\tau'), v(\tau'))}}=t_0 +\int_{\tau_0}^\tau\frac{2
   \sqrt{3}}{\sqrt{F(\tau' )}} d\tau' \nonumber \\
	&+ \int_{\tau_0}^\tau \frac{2 f_1 \mu 
   \text{csch}(\Delta (\tau' ))
   \text{sech}(\Delta (\tau' ))
   (U_0-F(\tau' )) \left(2
   \sqrt{3} \left(\gamma
   ^2-3\right) \cosh (2 \Delta
   (\tau' ))-3 \sqrt{2} \gamma
   \right)}{\sqrt{\sqrt{6}-\gamma }
   \left(\gamma
   +\sqrt{6}\right)^{3/2} F(\tau'
   )^{3/2}}d\tau' \nonumber\\
	& +\int_{\tau_0}^\tau\frac{6
   \left(\sqrt{3}-\sqrt{2} \gamma
   \right) f_2 \mu  (F(\tau'
   )-U_0)}{F(\tau'
   )^{3/2}} d\tau' + \int_{\tau_0}^\tau \frac{6 \left(\sqrt{2}
   \gamma +\sqrt{3}\right) f_3 \mu
    (F(\tau' )-U_0)}{F(\tau
   )^{3/2}}  +O(\mu^2),
	\end{align}
	\end{subequations} which generalize solutions \eqref{solutions-induced-gravity}.
	
In \eqref{cosmol_sol} we have introduced the expressions
\begin{align*}
&K_1=e^{3 \sqrt{\frac{3}{2}}
   \gamma  \left(c_2-c_3\right)-3
   \left(c_2+c_3\right)},\\
&K_3={\left(\frac{c_1}{\gamma}\right)}^2 e^{2 \sqrt{6} \gamma 
   \left(c_2-c_3\right)-6 c_2-6
   c_3},\\
& \alpha_2=\frac{\sqrt{\sqrt{6}-\gamma
   } \gamma  \left(\sqrt{2} \gamma
   ^3+3 \sqrt{3} \gamma ^2-6
   \sqrt{3}\right)
   f_2}{\left(\gamma
   +\sqrt{6}\right)^{3/2}
   \left(\gamma ^2-6\right)},
	\\
& \alpha_3=-\frac{2
   \sqrt{\sqrt{6}-\gamma } \gamma 
   \left(\sqrt{2} \gamma ^3+5
   \sqrt{3} \gamma ^2+12 \sqrt{2}
   \gamma +6 \sqrt{3}\right)
   f_3}{\left(\gamma
   +\sqrt{6}\right)^{3/2}
   \left(\gamma ^2-6\right)},
	\end{align*}
	and
\begin{align*}
 & F(\tau)=	U_0+\left(2-3 \gamma ^2\right)
   K_3 \sinh ^{\frac{12}{\sqrt{6}
   \gamma +6}-4}(\Delta (\tau)) \cosh
   ^{-\frac{12}{\sqrt{6} \gamma
   -6}-4}(\Delta (\tau)).
	\end{align*}
We note that by fine-tuning the parameter values we may get some integrable
cosmological models in the case of a massless and slowly moving scalar field. This makes our choice of potential very interesting. This issue, and the discussion of massive scalar field, $\phi$, deserves further investigation and it is left to future projects. However, it is worth noticing that the main focus of this research is not to find analytical solutions but to study the asymptotic behavior of the solutions space without using fine-tuning of the parameters and the initial conditions. 
Dynamical systems theory is a powerful tool for doing this
research.

\subsection{Dynamical system analysis}\label{SECT:2.1}
In order to study the cosmological behavior in a general way, independently of the
initial conditions and the specific universe evolution, we apply the dynamical
systems method, which allows to extract the global features of a cosmological scenario
\cite{FerreiraJoyce1997, CopelandLiddleWands1998, WainwrightEllis2005, Perko2008, ChenGongSaridakis2009, GiamboMiritzis2010, Escobar2011, XuSaridakisLeon2012,  LeonFadragas2012, Escobar2012, Escobar2013, Fadragas2013, CotsakisKittou2013, Coley2013, Leon:2014bta, Fadragas2014}. In this procedure, one first transforms the involved
cosmological equations into an autonomous system and then one extracts its critical points. 
Hence, taking linear perturbations around these critical points, and expressing the perturbations 
in terms of a perturbation matrix, allows to determine the type and stability of each 
critical point by examining the eigenvalues of this matrix. In the case of non-hyperbolic critical points one should use the center manifold theorem \cite{WainwrightEllis2005, Perko2008, Wiggins2010, Escobar2011, LeonFadragas2012, Escobar2012, Escobar2013, Fadragas2013, Coley2013, AlhoHellUggla2015}. 

\subsubsection{Finite analysis}\label{FA}

Let us define the following dynamical variables 
\begin{equation}\label{vars}
	\epsilon= \sqrt{\Phi}, \quad {x}=\frac{\dot\phi}{\sqrt{6} {H} \sqrt{\Phi} }, \quad {y}=\frac{\dot\Phi}{\Phi  {H}}, \quad {z}=\frac{\sqrt{ {U}(\Phi)}}{\sqrt{3} {H} \sqrt{\Phi}}\equiv \frac{1}{H}\sqrt{\frac{U_0}{3}}\Phi^{\frac{\gamma-\lambda_U}{2\gamma}}.
\end{equation}

The Friedmann equation \eqref{Fried_JF} leads to  
\begin{equation}
\label{FRIED}
\frac{V(\phi)}{3 H^2 \Phi}+ {x}^2+ {y}\left( \frac{\omega_0}{6}{y} -1\right)+ {z
}^2=1. 
\end{equation}
It is defined the auxiliary variable \cite{Garcia-SalcedoGonzalezQuiros2015}:
\begin{equation}
\Omega_{K}^{\text{eff}}={y}\left( \frac{\omega_0}{6}{y} -1\right),
\end{equation}
that it is interpreted as the dimensionless kinetic energy density of the JBD field $\Phi,$ and it is not necessarily positive. This implies that $ {x}^2+ {z
}^2$  might be greater than the unity. This is due to $\Phi$ is a non-conventional scalar field that does not follow the standard energy conditions for a scalar field in GR. Besides,  physical conditions $\Phi\ge0$, $U_0\ge0$ and $H\ge0$ implies $z\geq 0$. 

The evolution equations for the variables \eqref{vars} are given by:  
\begin{subequations}
	\label{DS_Jordan}
	\begin{align}
	&  {x}'=\left(3-\frac{9 \gamma ^2}{2}\right) x^3+\lambda_V \epsilon  \left(-\sqrt{\frac{3}{2}}
	\left(x^2-1\right)+\frac{\left(3 \gamma ^2-2\right) y^2}{4 \sqrt{6} \gamma
		^2}+\sqrt{\frac{3}{2}} y-\sqrt{\frac{3}{2}} z^2\right)+\nonumber \\ & +3 \left(\gamma ^2-1\right)
	x+\frac{\left(3 \gamma ^4-5 \gamma ^2+2\right) x y^2}{4 \gamma ^2}+\frac{1}{2} \left(6 \gamma
	^2-5\right) x y+\frac{3}{2} \gamma  x z^2 (\lambda_U-2 \gamma ),
	\\
	&  {y}'=3 \gamma ^2 \left(2-3 x^2\right)+y \left(9 \gamma ^2+\left(3-\frac{9 \gamma ^2}{2}\right)
	x^2+\frac{3}{2} \gamma  z^2 (\lambda_U-2 \gamma )-3\right)+\nonumber \\ & +\frac{1}{4} \left(3
	\gamma ^2+\frac{2}{\gamma ^2}-5\right) y^3+\left(\frac{9 \gamma ^2}{2}-4\right) y^2+3 \gamma 
	z^2 (\lambda_U-2 \gamma ),
	\\
	& {z}'= z \left(3 \gamma ^2+\left(3-\frac{9 \gamma ^2}{2}\right) x^2\right)+\frac{1}{4} \left(3 \gamma
	^2+\frac{2}{\gamma ^2}-5\right) y^2 z+\nonumber \\ & -\frac{y z \left(-6 \gamma ^3+3 \gamma +\lambda_U\right)}{2 \gamma } +\frac{3}{2} \gamma  z^3 (\lambda_U-2 \gamma ),
	\\
	& \epsilon'=\frac{ {y} \epsilon }{2},	
	\end{align}
\end{subequations}
where the comma denotes derivative with respect the conformal time $\tau=\ln  {a}$.

From the equations \eqref{DS_Jordan} it follows that the signs of $\epsilon$ and $z$ are invariant in time. This means that all the solutions with $\epsilon(\tau_0)= 0$ (respectively $\epsilon(\tau_0)<0$ or $\epsilon(\tau_0)>0$) at an initial time $\tau_0$, will satisfy  $\epsilon(\tau)= 0$ (respectively $\epsilon(\tau)<0$ or $\epsilon(\tau)>0$) at any time $\tau$.   
Given that the sign of $\epsilon$ is invariant for the flow,  we can safely analyze the stability of the fixed points with $\epsilon=0$ in a neighborhood of it, containing both $\epsilon>0$ and $\epsilon<0$ points, but since the boundary $\epsilon=0$ cannot be crossed, we restrict ourselves to the region of physical interest. The same reasoning applies to $z$. The ``recipe" to deal with this kind of dynamical systems was given in the seminal work \cite{CopelandLiddleWands1998}.
Henceforth, we can focus our analysis on solutions with $\epsilon\geq 0$ and $z\geq 0$, and the equations \eqref{DS_Jordan} define a flow on the phase space:
\begin{equation}
\label{Phase_Space:J}
\Psi:=\left\{( {x}, {y}, {z},\epsilon)\in\mathbb{R}^4:  {x}^2+ {y}\left( \frac{\omega_0}{6}{y} -1\right)+ {z
}^2\leq 1, \epsilon\geq 0,  {z}\geq 0\right\}.
\end{equation}

Furthermore, the cosmological parameters are given by:
\begin{subequations}
	\label{JFObs1x}
	\begin{align}
	& {\Omega}_1\equiv \frac{{\rho}_1}{3H^2}=\frac{2 y^2 \epsilon ^2+3 \gamma ^2 \left(4-\epsilon ^2 \left((y+2)^2-4
		z^2\right)\right)}{12 \gamma ^2},\\
	& {\Omega}_2\equiv \frac{{\rho}_2}{3H^2}=\frac{\epsilon ^2 \left(3 \gamma ^2 \left((y+2)^2-4
		z^2\right)-2 y^2\right)}{8 \gamma ^2},
	\\
	& {q}\equiv -1-\frac{\dot H}{H^2}=-1+3 \gamma ^2+\left(3-\frac{9 \gamma ^2}{2}\right) x^2+\frac{1}{4} \left(3 \gamma
	^2+\frac{2}{\gamma ^2}-5\right) y^2+\nonumber\\ & +\left(3 \gamma ^2-2\right) y+\frac{3}{2} \gamma  z^2
	(\lambda_U-2 \gamma ), \label{JFObs1q}	
	\end{align}
	\begin{align}
	& {w}_{\text{tot}}\equiv -1-\frac{2\dot H}{3 H^2}= -1+2 \gamma ^2+\left(2-3 \gamma ^2\right) x^2+\frac{1}{6}
	\left(3 \gamma ^2+\frac{2}{\gamma ^2}-5\right) y^2+\nonumber\\ &+\left(2 \gamma ^2-\frac{4}{3}\right)
	y+\gamma  z^2 (\lambda_U-2 \gamma ).
	\end{align}
\end{subequations}
	
	\paragraph{Fixed points and stability in the Jordan frame.}\label{A3}
	
	The critical points of the system in the Jordan frame \eqref{DS_Jordan} are:
	\begin{enumerate} 
		\item $J_{1,2}: \left( {x}= 0,  {y}= \pm \frac{6 \gamma }{\sqrt{6} \mp 3 \gamma }, {z}= 0,\epsilon= 0\right)$. They always exist. The eigenvalues are $$\left\{-\frac{3\gamma}{3\gamma\mp \sqrt{6}}, -\frac{3\gamma}{3\gamma\mp \sqrt{6}}, \frac{6
			\left(\sqrt{6}\mp 2 \gamma \right)}{\sqrt{6}\mp 3 \gamma },\frac{3 \left(\sqrt{6}\mp {\lambda_U}\right)}{\sqrt{6}\mp 3 \gamma }\right\}.$$ $J_1$ is a sink for $\sqrt{\frac{2}{3}}<\gamma <\sqrt{\frac{3}{2}}, {\lambda_U}<\sqrt{6}$, a source for ${\lambda_U}<\sqrt{6}, 0<\gamma <\sqrt{\frac{2}{3}}$ or a saddle otherwise.  $J_2$ is always a saddle. 
		\item $J_3: \left(0,\frac{2 \gamma ^2}{1-\gamma ^2},0,0\right)$. It exists for $-\sqrt{\frac{3}{2}}\leq \gamma <-1,$ or  $-1<\gamma <1,$ or $1<\gamma \leq \sqrt{\frac{3}{2}}$. The eigenvalues are $$\left\{-\frac{\gamma ^2}{\gamma ^2-1},-3,\frac{3-2 \gamma ^2}{\gamma ^2-1},\frac{\gamma 
			({\lambda_U}-2 \gamma )}{\gamma ^2-1}\right\}.$$ It is always a saddle.	
		\item $J_4: \left(0, 0, \frac{\sqrt{2} \sqrt{\gamma }}{\sqrt{2 \gamma -{\lambda_U}}}, 0\right)$. It exists for $\gamma\geq 0,\lambda_U\leq 0$. The eigenvalues are $$\left\{0,-3,\frac{1}{2} \left(-3-\sqrt{24 \gamma  {\lambda_U}+9}\right),\frac{1}{2}
		\left(-3+\sqrt{24 \gamma  {\lambda_U}+9}\right)\right\}.$$ It is nonhyperbolic with a 3D stable manifold for $\lambda_U<0, \gamma>0.$ Thus, it has a large probability to attract the universe at late times. The full stability analysis requires the application of the center manifold theorem (the analysis is done in subsection \ref{CENTER_J4_bd}). 
		\item $J_{5,6}: \left(\pm \sqrt{\frac{2}{3}},0,0,\pm \frac{2}{\lambda_U}\right)$. They always exist. 
		The eigenvalues are $$\left\{-1,2,\frac{1}{2} \left(-\sqrt{1-24 \gamma ^2}-1\right),\frac{1}{2} \left(\sqrt{1-24
			\gamma ^2}-1\right)\right\}.$$ Thus, they are always saddle. 
		\item $J_7: \left(0,-\frac{2 \gamma  {\lambda_U}}{\gamma  {\lambda_U}-2},\sqrt{\frac{2}{3}} \sqrt{\frac{6-{\lambda_U}^2}{(\gamma  {\lambda_U}-2)^2}}, 0\right).$ It exists for $-\frac{3}{2}<\omega_0<0, -\sqrt{6}\leq \lambda_U<\sqrt{2} \sqrt{2 \omega_0+3}$ or $-\frac{3}{2}<\omega_0<0, \sqrt{2} \sqrt{2 \omega_0+3}<\lambda_U\leq \sqrt{6}$ or $\omega_0=0, -\sqrt{6}\leq \lambda_U<\sqrt{6}$ or $\omega_0>0, -\sqrt{6}\leq \lambda_U\leq \sqrt{6}$.  The eigenvalues are $$\left\{\frac{\gamma  {\lambda_U}}{2-\gamma  {\lambda_U}},\frac{6-{\lambda_U} (\gamma +{\lambda_U})}{\gamma  {\lambda_U}-2},\frac{2 {\lambda_U} (2
			\gamma -{\lambda_U})}{\gamma  {\lambda_U}-2},\frac{6-{\lambda_U}^2}{\gamma  {\lambda_U}-2}\right\}.$$ It is always a saddle. 
		\item $J_{8,9}: \left(\pm\frac{\sqrt{{\lambda_U} (\gamma +{\lambda_U})-6}}{\sqrt{({\lambda_U}-2 \gamma )^2}},\frac{6 \gamma }{{\lambda_U}-2
			\gamma }, \frac{\sqrt{\gamma } \sqrt{\gamma +{\lambda_U}}}{\sqrt{({\lambda_U}-2 \gamma )^2}}, 0\right).$  They exist for \\ $-\frac{3}{2}<\omega_0<-\frac{5}{6}, \sqrt{\frac{12 \omega_0+\sqrt{24 \omega_0+37}+19}{2\omega_0+3}}\leq \lambda_U\leq \sqrt{6}$, or \\
		$\omega_0=-\frac{5}{6}, \frac{1}{4} \left(\sqrt{102}-\sqrt{6}\right)\leq \lambda_U<\sqrt{6}$, or $-\frac{5}{6}<\omega_0<-\frac{1}{2}, \sqrt{\frac{12 \omega_0+\sqrt{24 \omega_0+37}+19}{2 \omega_0+3}}\leq \lambda_U<\frac{2 \sqrt{2}}{\sqrt{2 \omega_0+3}},$  or $-\frac{5}{6}<\omega_0<-\frac{1}{2}, \frac{2 \sqrt{2}}{\sqrt{2\omega_0+3}}<\lambda_U\leq \sqrt{6}$, or $\omega_0=-\frac{1}{2}, 2<
		\lambda_U\leq \sqrt{6}$ or $\omega_0>-\frac{1}{2}, \sqrt{\frac{12 \omega_0+\sqrt{24 \omega_0+37}+19}{2\omega_0+3}}\leq \lambda_U\leq \sqrt{6}$. The eigenvalues are $$\left\{\frac{3 \gamma }{{\lambda_U}-2 \gamma },\frac{\mu_1}{2 \gamma 
			({\lambda_U}-2 \gamma )^3},\frac{\mu_2}{2 \gamma  ({\lambda_U}-2
			\gamma )^3},\frac{\mu_3}{2 \gamma  ({\lambda_U}-2 \gamma )^3}\right\},$$ where $\mu_1, \mu_2, \mu_3$ are the roots of $P(\mu)=-144 \gamma ^4 (\gamma +{\lambda_U}) ({\lambda_U} (\gamma +{\lambda_U})-6)
		(2 \gamma -{\lambda_U})^7+12 \gamma ^3 \mu  ({\lambda_U}-2 \gamma )^4
		\left(\left(\gamma ^2-12\right) {\lambda_U}+2 \gamma  {\lambda_U}^2+6 \gamma
		+{\lambda_U}^3\right)-6 \gamma  \mu ^2 (5 \gamma -2 {\lambda_U})
		({\lambda_U}-2 \gamma )^2+\mu ^3.$ Their stability should be analyzed numerically.
	\end{enumerate}
	
The existence and stability conditions of the above critical points 
 are displayed in Table \ref{Tab2}, whereas the cosmological parameters \eqref{JFObs1x} evaluated at the critical points, and the description of these critical points are given in Table \ref{TabBD1b}. 
\begin{landscape}
{\begin{table*}[ht!]
			\begin{tabular}{@{}c@{}c@{}c@{}c@{}}
				\hline \rule[-2ex]{0pt}{5.5ex} Label & $ {x},  {y},  {z}, \epsilon$ & Existence  & Stability  \\ 
				\hline \rule[-2ex]{0pt}{5.5ex} $J_{1,2}$ & $\left( 0,  \pm \frac{6 \gamma }{\sqrt{6} \mp 3 \gamma }, 0, 0\right)$  & always &  $J_1$ is a sink for $-\frac{5}{6}<\omega_0<0, {\lambda_U}<\sqrt{6}$ 		 	  \\ 
				&&& $J_1$ is a source for ${\lambda_U}<\sqrt{6},\omega_0>0$ \\
				&&& $J_1$ is a saddle otherwise   \\
				&&& $J_2$ is always a saddle   \\
				\hline \rule[-2ex]{0pt}{5.5ex}  $J_3$ & $\left(0,\frac{2 \gamma ^2}{1-\gamma ^2},0,0\right)$  & $\omega_0>-\frac{1}{2},$ or $-\frac{5}{6}<\omega_0<-\frac{1}{2}$  & saddle \\
				\hline \rule[-2ex]{0pt}{5.5ex}   $J_4$ & $\left(0, 0, \frac{\sqrt{2} \sqrt{\gamma }}{\sqrt{2 \gamma -{\lambda_U}}}, 0\right)$  & $\omega_0>-\frac{3}{2},\lambda_U\leq 0$ & sink for  $\omega_0>-\frac{3}{2}, \lambda_V\neq 0, \lambda_U<0$.  \\
				\hline \rule[-2ex]{0pt}{5.5ex}  $J_{5,6}$ & $\left(\pm \sqrt{\frac{2}{3}},0,0,\pm \frac{2}{\lambda_U}\right)$  & {$\lambda_U>0\ \text{for }J_5\ \text{and}\ \lambda_U<0\ \text{for }J_6$} & saddle \\ 
				\hline \rule[-2ex]{0pt}{5.5ex}  $J_7$ & $\left(0,-\frac{2 \gamma  {\lambda_U}}{\gamma  {\lambda_U}-2},\sqrt{\frac{2}{3}} \sqrt{\frac{6-{\lambda_U}^2}{(\gamma  {\lambda_U}-2)^2}}, 0\right)$  & $-\sqrt{6}\leq {\lambda_U}\leq \sqrt{6}, \gamma >0, \gamma\lambda_U\neq 2$ & saddle  \\ 
				\hline \rule[-2ex]{0pt}{5.5ex}  $J_{8,9}$ & $\left(\pm\frac{\sqrt{{\lambda_U} (\gamma +{\lambda_U})-6}}{\sqrt{({\lambda_U}-2 \gamma )^2}},\frac{6 \gamma }{{\lambda_U}-2
					\gamma }, \frac{\sqrt{\gamma } \sqrt{\gamma +{\lambda_U}}}{\sqrt{({\lambda_U}-2 \gamma )^2}}, 0\right)$  & $0<{\lambda_U}\leq \sqrt{6}, \gamma \geq \frac{6-\lambda_U^2}{{\lambda_U}}, \lambda_U\neq 2\gamma$ or & numerical inspection\\
				& & $\gamma \geq 0, \lambda_U>\sqrt{6}, \lambda_U\neq 2\gamma$ &  \\ 
				\hline 
			\end{tabular}  
			\caption{\label{Tab2} The existence and stability conditions of the critical points of 
				\eqref{DS_Jordan}. We use the definition $\gamma=(\omega_0+3/2)^{-1/2}.$}
		\end{table*}}
	\end{landscape}

	\begin{landscape}

		\begin{table*}[ht!]
\centering
\resizebox{\linewidth}{!}{

				\begin{tabular}{|@{ }c@{ }|c@{ }|c@{ }|c@{ }|c@{ }|c@{ }|c@{}|c@{}|}
					\hline \rule[-2ex]{0pt}{5.5ex} Label & $\Omega_1$ & $\Omega_2$ & $\Omega_{K}^{\text{eff}}$ & $q$ & $w_{\text{tot}}$  & Description &  $H(t)$ \\ 
					\hline \rule[-2ex]{0pt}{5.5ex} $J_{1,2}$   & $1$& $0$ & $1$ & $\frac{2 \sqrt{6}}{\sqrt{6}\mp 3 \gamma }$   & $\frac{\gamma \pm \sqrt{6}}{\sqrt{6}\mp 3 \gamma
					}$ &    $J_{1,2}$ are dominated by  $\Omega_{K}^{\text{eff}}$. & $\frac{H_0}{1+3 H_0 \left(\frac{\sqrt{6}\mp \gamma}{\sqrt{6}\mp 3 \gamma}\right)(t-t_0)}.$ \\
					\rule[-2ex]{0pt}{5.5ex} &&&&&& $J_1$ is accelerating for	$-\frac{3}{2}<\omega_0<0.$ &\\
					\rule[-2ex]{0pt}{5.5ex} &&&&&& $J_2$ is always decelerating.&\\ 
					\hline \rule[-2ex]{0pt}{5.5ex} $J_3$  & $1$ & $0$ & $\frac{\gamma ^2 \left(3 \gamma ^2-4\right)}{3 \left(1-\gamma ^2\right)^2}$  & $-\frac{1-2 \gamma ^2}{1-\gamma ^2}$ & $-\frac{3-5 \gamma ^2}{3 \left(1-\gamma
						^2\right)}$ &  Scaling between  $\Omega_{K}^{\text{eff}}$ and $\frac{V(\phi)}{3 \Phi H^2}$ for $\omega_0\neq -\frac{5}{6}$, & $\frac{H_0}{1+\frac{H_0\gamma^2}{1-\gamma^2} (t-t_0)}.$ \\  
					\rule[-2ex]{0pt}{5.5ex} &&&&&& Dominated by  $\Omega_{K}^{\text{eff}}$ for  $\omega_0=-\frac{5}{6}$, & \\ 
					\rule[-2ex]{0pt}{5.5ex} &&&&&& Accelerating for $\omega_0>\frac{1}{2}$ or $-\frac{3}{2}<\omega_0<-\frac{1}{2}$. &  \\   
					\hline \rule[-2ex]{0pt}{5.5ex} $J_4$ & $1$ & $0$ & $0$ & $-1$ & $-1$  & Intermediate accelerated $a(t)\simeq e^{\alpha_1  t^{p_1}}$, $\alpha_1>0, 0<p_1<1.$ & $\simeq \alpha_1 p_1 t^{p_1-1}.$\\ 
					\hline \rule[-2ex]{0pt}{5.5ex} $J_{5,6}$ & $1-\frac{4}{\lambda_V^2}$&$\frac{4}{\lambda_V^2}$ & $0$ &$1$&$\frac{1}{3}$ &  Radiation dominated. & $\frac{H_0}{1+2 H_0 (t-t_0)}.$\\ 
					\hline 
					\rule[-2ex]{0pt}{5.5ex}  $J_7$ & $1$ & $0$ & $-\frac{\lambda_U (-3 \gamma  (\gamma  \lambda_U-4)-2 \lambda_U)}{3
						(\gamma  \lambda_U-2)^2}$  & $\frac{2-\lambda_U^2}{\gamma  \lambda_U-2}$ & $\frac{6-\lambda_U (\gamma +2 \lambda_U)}{3 \gamma  \lambda_U-6}$ & Dominated by the energy density of $\Phi$. & $\frac{H_0}{1+\frac{H_0 \lambda_U(\gamma-\lambda_U)}{\gamma\lambda_U-2}(t-t_0)}.$\\ 
					\hline \rule[-2ex]{0pt}{5.5ex}  $J_{8,9}$ & $1$ & $0$ & $\frac{3 \left(\gamma ^2-2 \gamma  \lambda_U+2\right)}{(\lambda_U-2 \gamma
						)^2}$ & $\frac{\gamma -2 \lambda_U}{2 \gamma -\lambda_U}$& $-\frac{\lambda_U}{2 \gamma -\lambda_U}$ & Scaling solution. & $\frac{H_0}{1+3 H_0\left(\frac{\gamma-\lambda_U}{2\gamma-\lambda_U}\right) (t-t_0)}$.\\  
					\hline 
				\end{tabular} 
				
				}
				\caption{\label{TabBD1b} Description of the cosmological parameters \eqref{JFObs1x} of the critical points of 
					\eqref{DS_Jordan}.}

		\end{table*}
	\end{landscape}

\paragraph{Center manifold analysis for the intermediate accelerated solution $J_4$.}\label{CENTER_J4_bd}
	
	In order to investigate the stability of the center manifold for $J_4$ we introduce the new variables
	\begin{subequations}
		\label{centerA}
		\begin{align}
		&u=\epsilon,\\
		&v_1=\frac{\lambda_U \lambda_V \epsilon }{\sqrt{6} (2 \gamma
			-\lambda_U)}+x,\\
		&v_2=\frac{\sqrt{\gamma } \left(12 \gamma ^2-\sqrt{24 \gamma 
				\lambda_U+9}-3\right)}{\sqrt{12 \gamma -6 \lambda_U} \sqrt{8 \gamma 
				\lambda_U+3}}+\frac{y \left(6 \gamma ^3-3 \gamma -\lambda_U\right)}{\sqrt{6} \sqrt{\gamma } \sqrt{2 \gamma -\lambda_U} \sqrt{8 \gamma 
				\lambda_U+3}}+\nonumber\\ & +\frac{z \left(-12 \gamma ^2+\sqrt{24 \gamma  \lambda_U+9}+3\right)}{2 \sqrt{24 \gamma  \lambda_U+9}},\\
		&v_3=-\frac{\sqrt{\gamma } \left(12
			\gamma ^2+\sqrt{24 \gamma  \lambda_U+9}-3\right)}{\sqrt{12 \gamma -6 \lambda_U} \sqrt{8 \gamma  \lambda_U+3}}+\frac{y \left(-6 \gamma ^3+3 \gamma
			+\lambda_U\right)}{\sqrt{6} \sqrt{\gamma } \sqrt{2 \gamma -\lambda_U}
			\sqrt{8 \gamma  \lambda_U+3}}+\nonumber\\ & +\frac{z \left(12 \gamma ^2+\sqrt{24 \gamma 
				\lambda_U+9}-3\right)}{2 \sqrt{24 \gamma  \lambda_U+9}},
		\end{align}
	\end{subequations}
	to obtain
	{{\small
			\begin{align}
			\left(
			\begin{array}{c}
			u' \\
			v_1' \\
			v_2' \\
			v_3'
			\end{array}
			\right)= \left(
			\begin{array}{cccc}
			0 & 0 & 0 & 0 \\
			0 & -3 & 0 & 0 \\
			0 & 0 & -\frac{1}{2} \left(3-\sqrt{24 \gamma  \lambda_{U}+9}\right) & 0 \\
			0 & 0 & 0 & -\frac{1}{2} \left(3+\sqrt{24 \gamma  \lambda_{U}+9}\right) \\
			\end{array}
			\right) \left(
			\begin{array}{c}
			u \\
			v_1 \\
			v_2 \\
			v_3
			\end{array}
			\right) + \left(
			\begin{array}{c}
			f \\
			g_1 \\
			g_2 \\
			g_3
			\end{array}
			\right)
			\end{align}
		}} 
		where $(f,g_1,g_2, g_3)^T$ is a vector of higher order terms. 
		
		Since the center subspace of the origin is tangent to the $\epsilon$-axis, it follows that the center manifold of the origin is given locally by the graph
		\begin{align}
		&\Big\{(u, v_1,v_2,v_3): v_1=h_1(u), v_2=h_2(u), v_3=h_3(u), \nonumber \\ &  h_1(0)=h_2(0)=h_3(0)=0, \nonumber \\ & h_1'(0)=h_2'(0)=h_3'(0)=0, |u|<\delta \Big\},
		\end{align}
		where $\delta$ is a positive small enough constant. 
		The functions $h_i, i=1,2,3$ satisfy a set of quasilinear ordinary differential equations which can be expressed symbolically as
		\begin{equation}
		\label{h's}
		\Big[h_i'(u)u'-v_i'\Big]\Big|_{v_i=h_i(u)}=0, i=1,2,3,
		\end{equation}
		where one must substitute $u', v_1', v_2', v_3'$ through \eqref{centerA} and use the replacement $v_1\rightarrow h_1(u), v_2\rightarrow h_2(u)$
		and $v_3\rightarrow h_3(u)$.
		
		Setting
		\begin{subequations}
			\begin{align}
			& h_1(u)=a_{11} u^2 + a_{12} u^3 +\mathcal{O}\left(u\right)^4, \\ & h_2(u)=a_{2 1} u^2 + a_{2 2} u^3 +\mathcal{O}\left(u\right)^4, \\ & h_3(u)=a_{3 1} u^2 + a_{3 2} u^3 +\mathcal{O}\left(u\right)^4, 
			\end{align}
		\end{subequations}
		in \eqref{h's}, equating to zero all the coefficients of equal powers of $u$, and solving for the $a_{i j}$'s we get up to fourth order:  
		{\small{
				\begin{subequations}
					\begin{align}
					& a_{1 1}= 0, a_{1 2}= \frac{\lambda_U \lambda_V^3 (\gamma  \lambda_U-3)}{3 \sqrt{6} (\lambda_U-2 \gamma )^3},\\
					&    a_{2 1}= \frac{\sqrt{\gamma } \lambda_U^2 \lambda_V^2 \left(\gamma  \left(3 \gamma  \sqrt{8 \gamma  \lambda_U+3}+5 \sqrt{3} \gamma 
						-2 \sqrt{3} \lambda_U\right)-2 \left(\sqrt{8 \gamma 
							\lambda_U+3}+\sqrt{3}\right)\right)}{2 \sqrt{2} (2 \gamma -\lambda_U)^{5/2}
						\left(8 \sqrt{3} \gamma  \lambda_U-3 \sqrt{8 \gamma  \lambda_U+3}+3
						\sqrt{3}\right)},a_{2 2}= 0,\\
					& a_{3 1}= \frac{\sqrt{\gamma } \lambda_U^2 \lambda_V^2
						\left(\gamma  \left(-3 \gamma  \sqrt{8 \gamma  \lambda_U+3}+5 \sqrt{3} \gamma -2
						\sqrt{3} \lambda_U\right)+2 \sqrt{8 \gamma  \lambda_U+3}-2
						\sqrt{3}\right)}{2 \sqrt{2} (2 \gamma -\lambda_U)^{5/2} \left(8 \sqrt{3} \gamma 
						\lambda_U+3 \left(\sqrt{8 \gamma  \lambda_U+3}+\sqrt{3}\right)\right)},
					a_{3 2}=
					0.
					\end{align}
				\end{subequations}
			}}
			Henceforth, the dynamics on the center manifold is given by 
			\begin{equation}
			\label{CENTERA}
			u'=\frac{\gamma  \lambda_U \lambda_V^2 u^3}{2 (\lambda_U-2 \gamma
				)^2}+\mathcal{O}\left(u\right)^5.
			\end{equation}
			Neglecting the fifth-order terms and integrating we find that 
			\begin{equation}
			\label{u_expansion}
			u(\tau)=\pm \frac{\sqrt{-(\lambda_U-2 \gamma )^2}}{\sqrt{2 c_1 (\lambda_U-2 \gamma
					)^2+\gamma  \lambda_U \lambda_V^2 \tau }},
			\end{equation}
			where $c_1$ is an integration constant that must be negative in order for $u$ to be real-valued. 
			Thus, for $\gamma>0, \lambda_U\notin\{0, 2\gamma\}, \lambda_V\neq 0$,  it follows that the origin, and then $J_4$, is stable provided $\lambda_U<0$. 
			
\paragraph{Special case: $\lambda_U=0$, $U(\Phi)\propto \Phi^2$.} 

We introduce the new variables
			\begin{subequations}
				\begin{align}
				& u_1=\epsilon, \\
				& u_2=\frac{1}{2} \left(2 \gamma ^2-1\right) (y-2 z+2),\\
				& v_1= x,\\
				& v_2=\frac{y}{2}-\gamma ^2 (y-2
				z+2). 
				\end{align}
			\end{subequations}
			{{\small
					\begin{align}
					\label{centerAspecial}
					\left(
					\begin{array}{c}
					u_1' \\
					u_2' \\
					v_1' \\
					v_2'
					\end{array}
					\right)= \left(
					\begin{array}{cccc}
					0 & 0 & 0 & 0 \\
					0 & 0 & 0 & 0 \\
					0 & 0 & -3 & 0 \\
					0 & 0 & 0 & -3 \\
					\end{array}
					\right) \left(
					\begin{array}{c}
					u_1 \\
					u_2 \\
					v_1 \\
					v_2
					\end{array}
					\right) + \text{higher order terms}.
					\end{align}
				}}

				Since the center subspace of the origin is tangent to the plane $u_1$-$u_2$, it follows that the center manifold of the origin is given locally by the graph
				\begin{align}
				&\Big\{(u_1, u_2, v_1,v_2): v_1=h_1(u_1,u_2), v_2=h_2(u_1,u_2), \nonumber \\ &  h_1(0,0)=h_2(0,0)=0, \mathbf{Dh}(0,0)=\mathbf{0}, u_1^2+u_2^2<\delta \Big\},
				\end{align}
				where $\mathbf{Dh}$ is the matrix of derivatives and $\delta$ is a positive small enough constant. 
				The functions $h_1, h_2$ satisfy a set of quasilinear partial differential equations which can be expressed symbolically as
				\begin{equation}
				\label{specialh's}
				\Big[\frac{\partial{h_i(u)}}{\partial{u_1}}u_1'+\frac{\partial{h_i(u)}}{\partial{u_2}}u_2'-v_i'\Big]\Big|_{v_i=h_i(u)}=0, i=1,2,
				\end{equation}
				where one must substitute $u_1', u_2', v_1', v_2',$ through \eqref{centerAspecial} and use the replacement $v_1\rightarrow h_1(u_1,u_2), v_2\rightarrow h_2(u_1,u_2)$.
				
				Setting 
				\begin{subequations}
					\begin{align}
					& h_1=a_{11} u_1^2 + a_{12} u_1 u_2 +a_{22} u_2^2 +\mathcal{O}(3),\\
					& h_2=b_{11} u_1^2 + b_{12} u_1 u_2 +b_{22} u_2^2 +\mathcal{O}(3),
					\end{align}
				\end{subequations}
				and plugging back in  \eqref{specialh's}, equating to zero all the coefficients of equal powers of $u_1$ and $u_2$, and solving for the $a_{i j}$'s and $b_{i j}$'s we get up to third order
				\begin{equation}
				\left(
				\begin{array}{ccc}
				a_{11} & a_{12} & a_{22} \\
				b_{11} & b_{12} & b_{22}\\
				\end{array}
				\right)=\left(
				\begin{array}{ccc}
				0 & \frac{\sqrt{\frac{2}{3}} \lambda_V}{2 \gamma ^2-1} & 0 \\
				0 & 0 & \frac{\gamma ^2 \left(8 \gamma ^2-3\right)}{3 \left(1-2 \gamma ^2\right)^2} \\
				\end{array}
				\right).
				\end{equation}
				Thus, the dynamics on the center manifold is dictated by 
				\begin{subequations}
					\label{specialA_center}
					\begin{align}
					& u_1'=-\frac{2 \gamma ^2 u_1 u_2}{1-2 \gamma ^2} +\mathcal{O}(3),\\
					& u_2'=\frac{8 \gamma ^2 u_2^2}{1-2
						\gamma ^2} +\mathcal{O}(3),
					\end{align}
				\end{subequations}
				where $\mathcal{O}(3)$ denotes error terms of third order in the vector norm. 
				Neglecting the error terms and integrating out the system \eqref{specialA_center}
				we obtain
				\begin{equation}
				u_1=c_2 \sqrt[4]{\gamma ^2 \left(8 \tau -2 c_1\right)+c_1}, u_2=\frac{2 \gamma ^2-1}{\gamma ^2
					\left(8 \tau -2 c_1\right)+c_1}.
				\end{equation}
				Finally, it follows that for $\lambda_U=0$, the origin, and then the point $J_4$ behaves as a saddle since the orbits departs from the origin along the $\epsilon$-direction as the time goes forward. 
				
				\paragraph{Features of the critical points of the system \eqref{DS_Jordan}.}
				
				Let us summarize the features of the critical points of the system \eqref{DS_Jordan} found in subsection \ref{A3}:
\begin{enumerate}
	\item $J_{1,2}$ is dominated by the kinetic term of $\Phi$, that is $\Omega_K^{\text{eff}}=1$, and the quintessence field has $\Omega_2=0$. 
	$J_1$ is a sink for $-\frac{5}{6}<\omega_0<0$. This range for $\omega_0$ is several orders of magnitude lower than the bound $\omega_{0}>4\times 10^4$ imposed by the Solar System tests \cite{Will2014, BertottiIessTortora2003}, the bounds estimated on the basis of cosmological arguments $\omega_{0}>120$ \cite{AcquavivaBaccigalupiLeachEtAl2005} and $10<\omega_{0}< 10^7$ \cite{NagataChibaSugiyama2004}. Therefore, in section \eqref{JFDiscussion}  we will discuss on the asymptotics of $J_1$ when it is a source, i.e., for ${\lambda_U}<\sqrt{6},\omega_0>0$. $J_2$ is always a saddle. 
	
	\item $J_3$ is  a scaling solution between  $\Omega_{K}^{\text{eff}}$ and $\frac{V(\phi)}{3 \Phi H^2}$ for $\omega_0\neq -\sqrt{\frac{5}{3}}$. In this case,  $\frac{V(\phi)}{3 \Phi H^2}\rightarrow \frac{3-2\gamma^2}{2(1-\gamma^2)^2}$ and $\Phi\rightarrow 0$.   It is dominated by $\Omega_{K}^{\text{eff}}$ when  $\omega_0= -\sqrt{\frac{5}{3}}$,  and accelerating when $\omega_0>\frac{1}{2}$ or $-\frac{3}{2}<\omega_0<-\frac{1}{2}$. 
	
	\item $J_4$ represents an accelerating solution with $w_{\text{tot}}=-1$. At the end of section \ref{JFDiscussion} we will discuss on the corresponding asymptotics giving new arguments supporting the statements in        \cite{Garcia-SalcedoGonzalezQuiros2015} against \cite{HrycynaSzydlowski2013, HrycynaSzydlowski2013a, HrycynaSzydlowskiKamionka2014}. Indeed, $J_4$ represents an intermediate solution at late time, i.e., 
	$a(t)\simeq e^{\alpha_1  t^{p_1}}$ as $t\rightarrow \infty$
	where $\alpha_1>0,$ and $p_1:=-\frac{2 \gamma }{\lambda_U-3 \gamma }$, $0<p_1<1$ provided $\gamma>0, \lambda_U<0$. 
	Additionally,  $w_{\text{tot}}\rightarrow -1$ as  $t\rightarrow \infty$ for $\gamma>0, \lambda_V\neq 0, \lambda_U<0$.

	\item $J_{5,6}$ represent a radiation-dominated solution ($w_{\text{tot}}=\frac{1}{3}$) which are always saddle.
	
	\item $J_7$ denotes a solution dominated by the energy density of the scalar field $\Phi$. It is always a saddle. 
	
	\item $J_{8,9}$ represent scaling solutions where the contributions of $\Phi$ and $\phi$ to the total energy density are of the same order of magnitude.
	
\end{enumerate}

\paragraph{Hubble parameter of solutions near the fixed points.}\label{enumeration}

Now, we obtain first order approximation for the Hubble parameter of solutions near the fixed points of the system \eqref{DS_Jordan}, the results are presented in the last column of Table \ref{Tab2}. 
By definition we have $\dot H=-(1+q)H^2,$ with $q$ defined by \eqref{JFObs1q}. This expression is valid in the whole phase space and not just at the fixed points. In the case when $q\neq -1$, by continuity we can approximate the value of $q$ for a solution close to a given fixed point by the constant value $q=q^*\neq -1$, where the asterisk means evaluation at the fixed point. 

Defining the reference values $H_0=H|_{t=t_0}$, where $H_0$ and $t_0$ are finite numbers, we obtain 
	\begin{equation}\label{eqH}
	H(t)=\frac{H_0}{1+H_0(1+q^*) (t-t_0)},
	\end{equation}
which provides a first order approximation for the Hubble parameter at a given fixed point. 

These arguments are valid for all the fixed points, with the exception of $J_4$ since $q=-1$ for it.  For the point $J_4$ the above procedure can not be applied, instead we proceed as in Section \ref{JFDiscussion} to obtain $H\simeq \alpha_1 p_1 t^{p_1-1}$, with $0\leq p_1\leq 1$. Additionally, for the fixed points which are saddle this approximation is valid whenever the stable manifold of the fixed point is approached \footnote{For the solutions starting at the unstable manifold of the fixed point, the orbits depart from it, and the expression \eqref{eqH} might not be accurate.}.

From Table \ref{Tab2} we note that  the fixed points $J_4$, $J_7$, $J_{8}$ and $J_9$ have  $\epsilon=0$ (i.e., $\Phi=0$) and a finite non-zero value of $z\propto\Phi^{p}/H$, where $p= \frac{\gamma-\lambda_U}{\gamma}.$  This means that  $H$ and $\Phi^p$ ($p\neq0$) both tends to $\infty$ or to $0$, at the same rate, depending on the sign of $p$. That is, for $\gamma>\lambda_U$ we have $\Phi^p\rightarrow 0,  H\rightarrow 0$ and for $\gamma<\lambda_U$ we have $\Phi^p\rightarrow \infty,  H\rightarrow \infty$. On the other hand, from  Eq.\eqref{eqH} it follows that either  $H\rightarrow0$ as $t\rightarrow\infty$, or $H\rightarrow\infty$ for a finite value of time $t\rightarrow\frac{H_0t_0 (1+q^*)-1}{H_0 (1+q^*)}$, which corresponds to a finite time singularity (see references \cite{NojiriOdintsovTsujikawa2005, ElizaldeNojiriOdintsovEtAl2008} and the more recent work \cite{NojiriOdintsovOikonomou2015} for the classification of  finite-time future singularities).   Thus, we can combine the above facts to determine what happens with $H$ at $J_7$, $J_{8}$ and $J_9$, at least asymptotically.

\begin{enumerate}\label{enumeration}
\item For $J_7$ we have (for solutions starting at the stable manifold of $J_7$):
                  \begin{enumerate} 
									\item $-\sqrt{6}<\lambda_U\leq 0, \gamma >0$,  $H\rightarrow 0$.
									\item $0<\lambda_U\leq \sqrt{2}, 0<\gamma <\lambda_U$, $H\rightarrow \infty$ (at finite time). 
									\item $0<\lambda_U<\sqrt{2}, \lambda_U<\gamma <\frac{2}{\lambda_U}$,   $H\rightarrow 0$.
									\item $0<\lambda_U<\sqrt{2}, \gamma >\frac{2}{\lambda_U}$,   $H\rightarrow 0$.
									\item $\sqrt{2}<\lambda_U<\sqrt{6}, 0<\gamma <\frac{2}{\lambda_U}$, $H\rightarrow \infty$ (at finite time).
									\item $\sqrt{2}<\lambda_U<\sqrt{6}, \frac{2}{\lambda_U}<\gamma <\lambda_U$, $H\rightarrow \infty$ (at finite time).
									\item $\sqrt{2}\leq\lambda_U <\sqrt{6}, \gamma >\lambda_U$,   $H\rightarrow 0$.
									\end{enumerate}
\item For $J_{8,9}$ we have (for solutions starting at the stable manifold of $J_{8,9}$):
                  \begin{enumerate}
									\item $0<\lambda_U<\sqrt{3}, \gamma \geq \frac{6-\lambda_U^2}{\lambda_U}$, $H\rightarrow 0$.
									\item $\sqrt{3}<\lambda_U\leq 2, \frac{6-\lambda_U^2}{\lambda_U}\leq \gamma <\lambda_U$, $H\rightarrow \infty$ (at finite time).
									\item $2<\lambda_U<\sqrt{6}, \frac{6-\lambda_U^2}{\lambda_U}\leq \gamma <\frac{\lambda_U}{2}$, $H\rightarrow \infty$ (at finite time).
									\item $2<\lambda_U<\sqrt{6}, \frac{\lambda_U}{2}<\gamma <\lambda_U$, $H\rightarrow \infty$ (at finite time).
									\item $\sqrt{3}\leq \lambda_U\leq \sqrt{6}, \gamma >\lambda_U$, $H\rightarrow 0$.
									\item $\lambda_U>\sqrt{6}, 0\leq \gamma <\frac{\lambda_U}{2}$, $H\rightarrow \infty$ (at finite time). 
									\item $\lambda_U>\sqrt{6}, \frac{\lambda_U}{2}<\gamma <\lambda_U$, $H\rightarrow \infty$ (at finite time).
									\item $\lambda_U>\sqrt{6}, \gamma >\lambda_U$, $H\rightarrow 0$.
									\end{enumerate}									
\end{enumerate}
    
For $J_4$ \eqref{eqH} does not apply. However, since this point exists for $\gamma\ge0$ and $\lambda_U\le0$, and it is stable for $\gamma>0$, $\lambda_U<0$, then $p > 0$, thus, as $\Phi\rightarrow0$, $H\rightarrow0$ as the stable point is approached in order to $z$ be a non-zero constant at the fixed point, as verified in Sect. \ref{JFDiscussion}.

\subsubsection{Analysis at infinity}

As stated before $\Omega_{K}^{\text{eff}}$ is not necessarily positive, therefore, $ {x}^2+ {z
}^2$  might be greater than the unity, which implies that the phase space \eqref{Phase_Space:J} is not compact. So, in order to obtain global results about the dynamics, we implement a compactification scheme. For this analysis, the following compact variables are useful:
\begin{subequations}
	\label{POINCAREJBD}
	\begin{equation}
	X=\frac{x}{r}, \; Y=\frac{y}{r}, 
	\; Z=\frac{z}{r}, \; r= \sqrt{1+x^2+y^2+z^2},
	\end{equation}
	and 
	\begin{equation}
	E=\frac{\epsilon}{1+\epsilon},\label{varE}
		\end{equation}
\end{subequations}
which satisfy the evolution equations
\begin{subequations}
	\label{JF_Infinity}
	\begin{align}
	& \frac{d X}{d T}=-\frac{E \lambda_V (X^2-1) \left[12 \gamma ^2+12 \gamma ^2 K Y-24 \gamma ^2 X^2-9 \gamma ^2 Y^2-2 Y^2-24
		\gamma ^2 Z^2\right]}{4 \sqrt{6} \gamma ^2}+\nonumber \\  
	& +\frac{(E-1) K X \left[6 \gamma ^2 \left(\gamma ^2 \left(5 X^2+4 Z^2-2\right)-4 X^2+2\right)+\left(33 \gamma
		^4-7 \gamma ^2-2\right) Y^2\right]}{4 \gamma ^2}+ \nonumber\\ & -\frac{(E-1) \lambda_U X Z^2 \left[3 \gamma ^2 K-6 \gamma ^2 Y+Y\right]}{2 \gamma
	}+ \nonumber\\ & -\frac{1}{2} (E-1) X Y \left[3 \gamma ^2 \left(8 X^2+Y^2+6 Z^2-2\right)+5 X^2+8 Y^2+3 Z^2-5\right],
	\\	
	&\frac{d Y}{d T}= \frac{E \lambda_V X Y \left[3 \gamma ^2
		\left(-4 K Y+8 X^2+3 Y^2+8 Z^2-4\right)+2 Y^2\right]}{4 \sqrt{6} \gamma ^2}+ \nonumber\\ & +\frac{(E-1) K Y \left[6 \gamma ^2 \left(\gamma ^2 \left(5 X^2+4
		Z^2-6\right)-4 X^2+2\right)+\left(33 \gamma ^4-7 \gamma ^2-2\right) Y^2\right]}{4 \gamma ^2}+ \nonumber\\ & -\frac{(E-1) \lambda_U Z^2 \left[3 \gamma ^2 (Y
		(K-2 Y)+2)+Y^2\right]}{2 \gamma }+ \nonumber\\ & -\frac{1}{2} (E-1) \Big[Y^2 \left(5 X^2+8 Y^2+3 Z^2-8\right)+ \nonumber \\
	& +3 \gamma ^2 \left(2 X^2 \left(4 Y^2-5\right)+Y^4+Y^2
	\left(6 Z^2-5\right)-8 Z^2+4\right)\Big],
	\end{align}
	\begin{align}
	& \frac{d Z}{d T}=\frac{E \lambda_V X Z \left[3 \gamma ^2 \left(-4 K Y+8 X^2+3 Y^2+8 Z^2-4\right)+2 Y^2\right]}{4
		\sqrt{6} \gamma ^2}+ \nonumber\\ & +\frac{(E-1) K Z \left[-24 \gamma ^2 X^2+6 \gamma ^4 \left(5 X^2+4 Z^2-2\right)+\left(33 \gamma ^4-7 \gamma ^2-2\right)
		Y^2\right]}{4 \gamma ^2} + \nonumber\\ & +\frac{(E-1) \lambda_U Z \left[-3 \gamma ^2 Z^2 (K-2 Y)-Y Z^2+Y\right]}{2 \gamma }+ \nonumber\\ & -\frac{1}{2} (E-1) Y Z \left[3
	\gamma ^2 \left(8 X^2+Y^2+6 Z^2-2\right)+5 X^2+8 Y^2+3 Z^2-3\right],\\
	& \frac{d E}{d T}=\frac{1}{2} Y (E-1)^2 E,
	\end{align}
\end{subequations}
where $K=\sqrt{1-X^2-Y^2-Z^2}$, and we have introduced the new time variable $T$ given by $
{d T}\equiv (1-E)^{-1}K^{-1}{d\tau}$.
The physical region of the phase space becomes
\begin{align}
&\Psi_{\infty}:=\Big\{(X,Y,Z,E):  2 X^2+Y^2 \left(\frac{\omega_0}{6}+1\right)- Y \sqrt{1-X^2-Y^2-Z^2}+2 Z^2 \leq 1, \nonumber \\
& \qquad\qquad Z\geq 0, X^2+Y^2+Z^2\leq 1,  0\leq E \leq 1\Big\}.
\end{align}
The critical points of the system  \eqref{DS_Jordan} at infinity are the critical points of
\eqref{JF_Infinity} located on 
\begin{align}
&\Big\{(X,Y,Z,E):  2 X^2+Y^2 \left(\frac{\omega_0}{6}+1\right)- Y \sqrt{1-X^2-Y^2-Z^2}+2 Z^2 \leq 1, \nonumber \\
& \qquad\qquad Z\geq 0, X^2+Y^2+Z^2= 1,  0< E < 1\Big\}\cup \nonumber \\
& \qquad\qquad \Big\{(X,Y,Z,E):  2 X^2+Y^2 \left(\frac{\omega_0}{6}+1\right)- Y \sqrt{1-X^2-Y^2-Z^2}+2 Z^2 \leq 1, \nonumber \\
& \qquad\qquad Z\geq 0, X^2+Y^2+Z^2< 1,  E = 1\Big\}.
\end{align}

 {\begin{table*}
 		\begin{tabular}{@{}c@{}c@{}c@{}}
 			\hline \rule[-2ex]{0pt}{5.5ex} Label & $ X,  Y, Z, E$ & Existence  \\ 
 			\hline \rule[-2ex]{0pt}{5.5ex} $Q_{1,2}$ & $\left(0,\pm 1,0,0\right)$  & $-\frac{3}{2}\leq \omega_0\leq 0$   \\ 
 			\hline \rule[-2ex]{0pt}{5.5ex} $Q_{3,4}$ & $\left(\pm\sqrt{\frac{2 \omega_0+1}{2 \omega_0-11}},-\frac{2 \sqrt{3}}{\sqrt{11-2 \omega_0}},0,0\right)$  & $-\frac{3}{2}\leq \omega_0\leq -\frac{1}{2}$   \\ 
 			\hline \rule[-2ex]{0pt}{5.5ex} $Q_{5,6}$ & $\left(\pm\sqrt{\frac{2 \omega_0+1}{2 \omega_0-11}},\frac{2 \sqrt{3}}{\sqrt{11-2 \omega_0}},0,0\right)$  & $-\frac{3}{2}\leq \omega_0\leq -\frac{1}{2}$   \\
 			\hline \rule[-2ex]{0pt}{5.5ex} $Q_{7,8}$ & $\left(0,Y_{7,8}^*,\sqrt{1-{Y_{7,8}^*}^2},0\right)$ & $-\frac{3}{2}< \omega_0< -\frac{1}{2}, \frac{\sqrt{2} (10  \omega_0+9)}{(2  \omega_0+3)^{3/2}}\leq \lambda_U\leq \sqrt{4  \omega_0+6}$\\
 			\hline \rule[-2ex]{0pt}{5.5ex} $Q_{9,10}$ & $\left(\pm\frac{\sqrt{\omega_0}}{\sqrt{\omega_0-6}},-\frac{\sqrt{6}}{\sqrt{6-\omega_0}},0,1\right)$  & $-\frac{3}{2}\leq \omega_0\leq 0$   \\ 
 			\hline \rule[-2ex]{0pt}{5.5ex} $Q_{11,12}$ & $\left(\pm\frac{\sqrt{\omega_0}}{\sqrt{\omega_0-6}},\frac{\sqrt{6}}{\sqrt{6-\omega_0}},0,1\right)$  & $-\frac{3}{2}\leq \omega_0\leq 0$   \\
 			\hline \rule[-2ex]{0pt}{5.5ex} $Q_{13,14}$ & $\left(X_c,\pm \frac{\sqrt{6}}{\sqrt{6-\omega_0}},\frac{\sqrt{X_c^2 \omega_0-6 X_c^2-\omega_0}}{\sqrt{6-\omega_0}},1\right)$  & $-\frac{3}{2}\leq \omega_0\leq 0,  X_c^2\leq \frac{\omega_0}{\omega_0-6}$   \\
 			\hline  
 			\hline 
 		\end{tabular} 
 		\caption{\label{Tabinfinity} Existence conditions of the critical points at infinity of the system \eqref{DS_Jordan}, located at the sphere $X^2+Y^2+Z^2=1$. We use the notation $Y_{7,8}^*=\pm\frac{2 \sqrt{3} \sqrt{-2  \lambda_U  \omega_0-3  \lambda_U+2 \sqrt{2} \sqrt{2  \omega_0+3}}}{\sqrt{ \lambda_U
 					(2  \omega_0-9) (2  \omega_0+3)+5 \sqrt{4  \omega_0+6} (3-2  \omega_0)}}.$ $X_c$ is an arbitrary parameter.}
 	\end{table*}}
 
The critical points of the system \eqref{DS_Jordan} at infinity, located at the Poincar\'e sphere $X^2+Y^2+Z^2=1$, are shown in Table \ref{Tabinfinity}. None of these points can be considered of physical relevance  since their existence conditions are in stress with the observational bounds on $\omega_{0}$ obtained in \cite{Will2014, BertottiIessTortora2003, NagataChibaSugiyama2004}. 
On the other hand, in order to describe the critical points at infinity, which satisfy $E=1$  and $X^2+Y^2+Z^2< 1$, we choose: the variables $x,y,z$ defined in \eqref{vars}, the variable $E$ defined in \eqref{varE} and a new time variable $\check{T}$ defined by  ${d \check{T}}\equiv (1-E)^{-1}{d\tau}$. 
The resulting system admits the set of non-hyperbolic equilibrium points \footnote{All the eigenvalues of the Jacobian matrix are zero.} at infinity $z_c=\sqrt{1-x_c^2-\frac{y_c^2 \omega_0}{6}+y_c}, E=1,$ representing the boundary of the phase space $\Psi$ but with $\Phi\rightarrow \infty$.

\subsection{Viability of the intermediate accelerated solution in the Jordan Frame }\label{JFDiscussion}
The possible future attractors of our model, in the Jordan frame, are critical points $J_1$ and $J_4$.  
$J_{1}$ is an attractor for $-\frac{5}{6}<\omega_0<0$, but this existence condition is at variance with the observational bounds reported in \cite{Will2014, BertottiIessTortora2003, AcquavivaBaccigalupiLeachEtAl2005, NagataChibaSugiyama2004}, thus, the discussion of it is omitted \footnote{We restrict ourselves to positive values of BD parameter, $w_{0}>0$, despite of the observational results reported in \cite{Will2014, BertottiIessTortora2003, AcquavivaBaccigalupiLeachEtAl2005, NagataChibaSugiyama2004} could be less restrictive in the cases of modifed JBD theories.}.
 On the other hand,  $J_4$  is a late-time attractor for $\gamma>0, \lambda_V\neq 0, \lambda_U<0$, and the equation of state parameter becomes $w_{\text{tot}}=-1$ as the critical point is approached. Let us discuss more on the corresponding asymptotics.
 	
 	From the definition of $z$ it follows that at the equilibrium point 
 	\begin{equation}
 	\label{approxH_1}
 	H=\frac{\sqrt{(2\gamma-\lambda_U)U_0}}{\sqrt{6\gamma}} \Phi^{\frac{\gamma -\lambda_U}{2\gamma}}.
 	\end{equation}
Besides, under the conditions $\gamma>0, \lambda_V\neq 0, \lambda_U<0$ it follows that $H$ necessarily tends to zero when the critical point is approached  since $\Phi$ tends to zero at the critical point. Furthermore, since the deceleration parameter satisfies $q\rightarrow -1$ at the critical point, it follows by continuity that $\dot H\ll H^2$ at late times, which means that $\dot H$ tends to zero too. 
 	
 	Now, let us take advantage of the formula \eqref{u_expansion}, which is valid up to fifth order, for obtaining some asymptotic expansions.
 	Since $\Phi=u^2$, then we get
 	\begin{equation}
 	\label{approx1}
 	\Phi(a)=-\frac{(\lambda_U-2 \gamma )^2}{\gamma  \lambda_U \lambda_V^2 \ln
 		(a)+2 c_1 (\lambda_U-2 \gamma )^2}.
 	\end{equation}
 	Substituting \eqref{approx1} in \eqref{approxH_1} it follows 
 	\begin{equation}
 	\label{approxH_2}
 	H=\frac{\sqrt{U_0} \sqrt{\frac{2 \gamma -\lambda_U}{\gamma }} \left(-\frac{(\lambda_U-2 \gamma )^2}{\gamma  \lambda_U
 			\lambda_V^2 \ln (a)+2 c_1 (\lambda_U-2 \gamma )^2}\right){}^{\frac{\gamma -\lambda_U}{2 \gamma }}}{\sqrt{6}}.
 	\end{equation} Both expressions tend to zero as $a\rightarrow \infty$, provided $\gamma>0, \lambda_V\neq 0, \lambda_U<0$.
 	Integrating out \eqref{approxH_2} for $a$ it follows
 	{\small{\begin{align}
 			a&=\exp \left[-\frac{(\lambda_U-2 \gamma )^2 \left(2^{\frac{2 \gamma }{\lambda_U-3 \gamma }} \left(-\frac{\gamma  (\lambda_U-2
 					\gamma )^2 \left(\sqrt{6} t \sqrt{U_0} \sqrt{2-\frac{\lambda_U}{\gamma }}-6 c_2\right)}{\lambda_U \lambda_V^2 (3
 					\gamma -\lambda_U) \left(t^2 U_0 (2 \gamma -\lambda_U)-6 \gamma  c_2^2\right)}\right){}^{\frac{2 \gamma }{\lambda_U-3
 						\gamma }}+2 c_1\right)}{\gamma  \lambda_U \lambda_V^2}\right]\nonumber\\
 			& \simeq e^{\alpha_1  t^{p_1}} (\text{leading terms as t } \rightarrow \infty),
 			\end{align} }} where $c_1$ and $c_2$ are integration constants, and \\
 	$\alpha_1=-\frac{2^{\frac{2 \gamma }{\lambda_U-3 \gamma }} (\lambda_U-2 \gamma )^{\frac{2 \gamma }{\lambda_U-3 \gamma }+2}
 		\left(12-\frac{6 \lambda_U}{\gamma }\right)^{\frac{\gamma }{\lambda_U-3 \gamma }} \left(\lambda_U
 		\left(3-\frac{\lambda_U}{\gamma }\right)\right)^{-\frac{2 \gamma }{\lambda_U-3 \gamma }} \lambda_V^{-\frac{4 \gamma
 			}{\lambda_U-3 \gamma }-2} U_0^{\frac{\gamma }{3 \gamma -\lambda_U}}}{\gamma  \lambda_U},$ \\
  and $p_1=-\frac{2 \gamma }{\lambda_U-3 \gamma },$
 	where $\alpha_1>0,$ and $p_1 >0$, provided $\gamma>0, \lambda_U<0$. 
 	
 	The expressions for the deceleration parameter is 
 	{\small{\begin{align}
 			& q=-1+\frac{ 2^{-\frac{2 \gamma }{\lambda_U-3 \gamma }-1} \lambda_U \lambda_V^2 (\lambda_U-\gamma ) \left(-\frac{\gamma 
 					(\lambda_U-2 \gamma )^2 \left(\sqrt{6} t \sqrt{U_0} \sqrt{2-\frac{\lambda_U}{\gamma }}-6 c_2\right)}{\lambda_U
 					\lambda_V^2 (3 \gamma -\lambda_U) \left(t^2 U_0 (2 \gamma -\lambda_U)-6 \gamma  c_2^2\right)}\right){}^{-\frac{2
 						\gamma }{\lambda_U-3 \gamma }}}{(\lambda_U-2 \gamma )^2}\nonumber \\
 			& \simeq  -1 -\frac{(p_1-1) t^{-p_1}}{\alpha_1  p_1} \; (\text{leading terms as t } \rightarrow \infty).
 			\end{align}}}
 	
 	Let us observe that $a\rightarrow \infty$ as $t\rightarrow \infty$. 
 	Additionally, it is recovered the expected effective equation of state parameter, $w_{\text{tot}}=-1$ as  $t\rightarrow \infty$ for $\gamma>0, \lambda_V\neq 0, \lambda_U<0$. 
 	
 	From the equations \eqref{FRIED}, \eqref{approxH_1}, and \eqref{approx1}, we obtain the following relation, which is valid at the critical point
 	 	\begin{equation}
 	 	\label{POTENTIAL}
 	V(\phi)= -\frac{U_0 \lambda_U}{2\gamma}\Phi^{\frac{2 \gamma -\lambda_U}{\gamma}}=-\frac{ U_0 \lambda_U \left(-\frac{(\lambda_U-2 \gamma
 			)^2}{\gamma  \lambda_U \lambda_V^2 \ln (a)+2 c_1
 			(\lambda_U-2 \gamma )^2}\right)^{2-\frac{\lambda_U}{\gamma
 			}}}{2 \gamma }.
 	\end{equation}
 	
 	Therefore, using the equations \eqref{approx1}, \eqref{approxH_2}, and \eqref{POTENTIAL}, we can obtain asymptotic expressions for $\Phi, H$ and $\phi$ in terms of $t$ after the substitution of $a\simeq e^{\alpha_1  t^{p_1}}$, which are valid as $t\rightarrow \infty$ for $\gamma>0, \lambda_V\neq 0, \lambda_U<0$.

 	Now, in order for $J_4$ to be a de Sitter solution it is required that $p_1=1$, which implies $\lambda_U=\gamma$. But this would lead to a contradiction, since $\gamma$ is assumed to be positive, and $\lambda_U$ is negative in order for $J_4$ to be an attractor. In conclusion, in the modified JBD theory, the de Sitter solution is not a natural attractor in the Jordan frame. Hence, we offer new arguments supporting the statements given in  \cite{Garcia-SalcedoGonzalezQuiros2015} against the validity of the results presented in \cite{HrycynaSzydlowski2013, HrycynaSzydlowski2013a, HrycynaSzydlowskiKamionka2014}.

	Furthermore, since $J_4$ corresponds to $\Phi=0$, and since $\Phi$ plays the role of an effective Planck mass, this would imply that there is no gravity at $\Phi=0$. Thus, one must explicitly show that the solutions with a very small $\Phi$ are suitable to describe the late-time universe. Indeed, we have 
		\begin{align}
	&R=6\dot H +12 H^2\nonumber\\
	&= 6(1-q)H^2\nonumber\\
	&=6\left(1+\frac{a\ddot a}{{\dot a}^2}\right)\left(\frac{\dot a}{a}\right)^2\nonumber\\
	&\simeq 12 \alpha_1^2 p_1^2 t^{2 (p_1-1)}+6 \alpha_1 (p_1-1) p_1 t^{p_1-2},
	\end{align}
	where in the last line we evaluated the approximate solution  $a\simeq e^{\alpha_1  t^{p_1}}$ valid for $\gamma>0, \lambda_V\neq 0, \lambda_U<0$.  
		On the other hand, the leading terms as $a\rightarrow \infty$ in \eqref{approx1} are 
	\begin{equation}
	\Phi\simeq -\frac{(\lambda_U-2 \gamma )^2}{\gamma  \lambda_U \lambda_V^2 \ln
 		(a)}= -\frac{(\lambda_U-2 \gamma )^2}{\gamma  \lambda_U \lambda_V^2 \alpha_1 t^{p_1}}\sim \frac{1}{t^{p_1}}.
		\end{equation}
		Henceforth, the term $\Phi R$ in the action \eqref{Jordan_action} can be expressed as 
		\begin{equation}
		\Phi R \simeq A t^{p_1-2}+B t^{-2},
		\end{equation}
		where $A= -\frac{12 (\lambda_U-2 \gamma )^2\alpha_1 p_1^2}{\gamma  \lambda_U \lambda_V^2},$ $B=-\frac{6(\lambda_U-2 \gamma )^2 (p_1-1) p_1}{\gamma  \lambda_U \lambda_V^2},$ and  $p_1=-\frac{2 \gamma }{\lambda_U-3 \gamma }>0,$
$\alpha_1>0,$ provided $\gamma>0, \lambda_U<0$. Thus, $\Phi R$ tends to zero as the fixed point $J_4$ is approached, but it is always non-negative since $A$ and $B$ are both non-negative for $\gamma>0, \lambda_U<0$. Summarizing, we have explicitly shown that the solutions near $J_4$ satisfy $\Phi\sim \frac{1}{t^{p_1}}$,  $\Phi R \simeq A t^{p_1-2}+B t^{-2}$ and $H\simeq \alpha_1 p_1 t^{p_1-1}$, which tend asymptotically to zero but never reach this value. 
Besides, the effective gravitational coupling (the one measured in Cavendish-like experiments) \cite{Garcia-SalcedoGonzalezQuiros2015}, 
\begin{equation}
G_{\text{eff}}=\frac{4+2 \omega_0}{3+2 \omega_0}\Phi^{-1}
\end{equation} satisfies 
\begin{equation}
\frac{\dot G_{\text{eff}}}{G_{\text{eff}}}=-\frac{\dot \Phi}{\Phi}\simeq\frac{p_1}{t}.
\end{equation}
As a consequence of the above, if we consider cosmological constraints on the variability of the gravitational constant \cite{Uzan2003}, for instance the ones in \cite{WuChen2010}, which uses WMAP-5yr data combined with SDSS power spectrum data: $$-1.75\times 10^{-12}\,\text{yr}^{-1}<\frac{\dot G}{G}<1.05\times 10^{-12}\,\text{yr}^{-1},$$ or the ones derived in Ref. \cite{AccettaKraussRomanelli1990}, where the dependence of the abundances of the D, $^3$He, $^4$He, and $^7$Li upon the variation of $G$ was analyzed: $$|\dot G/G|<9\times 10^{-13}\,\text{yr}^{-1},$$ one see that for a given $p_1$ and a large $t$ the above constraints are very easily fulfilled. Hence, there are solutions with very small $\Phi$ which are suitable to describe the late-time universe.

Now, let us use the above arguments to obtain values of the free parameters that lead to values of $\frac{\dot G_{\text{eff}}}{G_{\text{eff}}}$ according to observations. 
Taking the Hubble time to be $t_0=13.817\times 10^9$ yr (as, for instance, in \cite{HrycynaSzydlowski2013a}), i.e., the present value of the Hubble constant $H_0=7.24\times 10^{-11}$ yr$^{-1}$, and given the value of $\lambda_U<0$, one gets  
$$\omega_0>\frac{18186.2}{\lambda_U^2}-1.5 \implies -1.75\times 10^{-12}\,\text{yr}^{-1}<\frac{\dot G_{\text{eff}}}{G_{\text{eff}}}<1.05\times 10^{-12}\,\text{yr}^{-1}$$
and 
$$\omega_0>\frac{24911.1}{\lambda_U^2}-1.5 \implies \Big|\frac{\dot G_{\text{eff}}}{G_{\text{eff}}}\Big|<9\times 10^{-13}\,\text{yr}^{-1}.$$

Conversely, given $\omega_0$, $$\lambda_U<-\frac{190.716}{\sqrt{2\omega_0+3}}  \implies -1.75\times 10^{-12}\,\text{yr}^{-1}<\frac{\dot G_{\text{eff}}}{G_{\text{eff}}}<1.05\times 10^{-12}\,\text{yr}^{-1}$$ and
$$\lambda_U<-\frac{223.209}{\sqrt{2 \omega_0+3}}\implies \Big|\frac{\dot G_{\text{eff}}}{G_{\text{eff}}}\Big|<9\times 10^{-13}\,\text{yr}^{-1}.$$

\subsection{The interplay with the induced gravity model}\label{interplay}

Recall that our model \eqref{Jordan_action} with ${U}(\Phi)=U_0 \Phi, \lambda_U=\gamma=\frac{1}{\sqrt{\omega_0+\frac{3}{2}}}, 0<\gamma<\sqrt{\frac{2}{3}}$ is equivalent to an extension of the so-called induced gravity model \cite{KamenshchikPozdeevaTronconiEtAl2014, AndrianovCannataKamenshchik2011} with action:

\begin{align}\label{Jordan_action_induced-gravity}
S_{JF}&=\int\sqrt{- {g}}\left(\frac{\Phi  {R}}{2}-\frac{2-3 \gamma ^2}{4 \gamma ^2 \Phi} {g}^{\mu\nu} {\partial}_{\mu}\Phi {\partial}_{\nu}\Phi- U_0 \Phi-\frac{1}{2} {g}^{\mu\nu} {\partial}_{\mu}\phi {\partial}_{\nu}\phi-V(\phi)\right)d^4x.
\end{align}

Let us observe that $J_4$ exists for $\lambda_U\leq 0, \gamma> 0$ and the condition for recovering the induced gravity model is $\gamma=\lambda_U, 0<\gamma<\sqrt{\frac{2}{3}}$. Thus, strictly speaking, $J_4$ does not exist for the induced gravity model.

In order to connect the main results of the section \eqref{FA} with the results in the section \eqref{massless_s_f}  we have to set $V(\phi)=0$ in \eqref{Jordan_action_induced-gravity}. Hence, the parameter space reduces to one parameter $0<\gamma<\sqrt{\frac{2}{3}}$, and the system \eqref{DS_Jordan} has the restriction
$${x}^2+ {y}\left[\left(\frac{1}{6\gamma^2}-\frac{1}{4}\right){y} -1\right]+ {z
}^2=1.$$  
This restriction allows to eliminate one variable, say $x$. Additionally, the equation for $\epsilon$ decouples.  Thus, it is obtained the reduced dynamical system
\begin{subequations}
\label{induced_grav_syst_A}
\begin{align}
& y'=\frac{1}{4} (y+2) y^2-\frac{3}{8} \gamma ^2
   (y+2) \left((y+2)^2-8 z^2\right)-3 y
   z^2,\\
& z'=	\frac{1}{8} \left(2 (y (y+4)+12) z-3 \gamma ^2
   (y+2)^2 z\right)+3 \left(\gamma ^2-1\right)
   z^3,
	\end{align}
\end{subequations}
defined on the invariant set 
\begin{align}
\left\{(y,z), {y}\left[\left(\frac{1}{6\gamma^2}-\frac{1}{4}\right){y} -1\right]+ {z
}^2\leq 1, z\geq 0\right\}.
\end{align}
The only critical points that exist for \eqref{induced_grav_syst_A}  at the finite region are those enumerated below.
\begin{enumerate}
\item $J_1$ always exists. The eigenvalues of the linearization of the reduced system \eqref{induced_grav_syst_A} are $\frac{6 \gamma }{\sqrt{6}-3 \gamma
   },\frac{3 \left(\sqrt{6}-\gamma
   \right)}{\sqrt{6}-3 \gamma }$. Thus, it is always the source. 
\item $J_2$ always exists. The eigenvalues of the linearization of the reduced system \eqref{induced_grav_syst_A} are $-\frac{6 \gamma }{3 \gamma
   +\sqrt{6}},\frac{3 \left(\gamma
   +\sqrt{6}\right)}{3 \gamma
   +\sqrt{6}}$. Thus, it is always a saddle. 
\item The point $J_7$ becomes $(y,z)= \left(\frac{2 \gamma ^2}{2-\gamma
   ^2},\frac{\sqrt{4-\frac{2 \gamma
   ^2}{3}}}{2-\gamma ^2}\right)$. The eigenvalues of the linearization of the reduced system \eqref{induced_grav_syst_A}  are now $-\frac{\gamma ^2-6}{\gamma
   ^2-2},-\frac{4 \left(\gamma^2-3\right)}{\gamma ^2-2}$. Thus, it is the sink in the plane $(y,z)$  for the range $0<\gamma<\sqrt{\frac{2}{3}}$. 
	\footnote{Although the point is unstable along the $\epsilon$-axis (then, a saddle) since $\epsilon'=\frac{ {y} \epsilon }{2}$ and $y>0$ at the fixed point. However, the analysis on the $\epsilon$-axis can be dropped out since the corresponding equation decouples.}  
\end{enumerate}

Concerning $J_7$, the results shown in the paragraph \eqref{enumeration} cannot be applied since $\lambda_U=\gamma$. Therefore, we introduce the new variable $\sigma=2\sqrt{\omega_0 \Phi}$ to get
\begin{subequations}
\label{exact}
\begin{align}
& y=\frac{2 \dot{\sigma}(t)}{H(t)
   \sigma (t)}, \\
&	z=\frac{\sqrt{U_0}}{\sqrt{3}  H(t)}. 
\end{align}
\end{subequations}

Now, we want to demonstrated that the solutions given by \eqref{cosmol_sol} converges to $J_7$ as $t\rightarrow \infty$. 

First of all, since the relative errors \eqref{relative-errors} can be made small for large enough values of time, then we can approximate \eqref{cosmol_sol} by 
	\begin{subequations}
	\label{solutions-induced-gravity_2}
	\begin{align}
	&\sigma(t)=e^{\sqrt{\frac{3}{2}} \gamma 
   \left(c_2-c_3\right)} \sinh
   ^{-\frac{\sqrt{6}\gamma}{\sqrt{6} \gamma
   +6}}(\Delta(t) ) \cosh
   ^{-\frac{\sqrt{6}\gamma}{\sqrt{6} \gamma
   -6}}(\Delta(t) ),\\
	&a(t)=e^{\sqrt{\frac{3}{2}} \gamma  \left(c_3-c_2\right)+c_2+c_3} \sinh ^{\frac{\sqrt{6} \gamma
   +2}{\sqrt{6} \gamma +6}}(\Delta (\tau
   )) \cosh ^{\frac{\sqrt{6} \gamma
   -2}{\sqrt{6} \gamma -6}}(\Delta (\tau
   ))
	,\\
	&H(t)=\frac{\sqrt{U_0
	} \text{csch}(2 \Delta (t)) \left(2 \sqrt{6} \gamma -3 \left(\gamma
   ^2-2\right) \cosh (2 \Delta (t))\right)}{3 \sqrt{2} \sqrt{6-\gamma ^2}},
	\end{align}
	\end{subequations}
where $\Delta(t)=\frac{\sqrt{6-\gamma ^2}
   \sqrt{U_0} \left(24
   c_1+t\right)}{2\sqrt{2}}.$

Secondly, substituting \eqref{solutions-induced-gravity_2} in \eqref{exact} we obtain 
\begin{subequations}
\begin{align}
	y=\frac
   {6 \gamma  \left(\sqrt{6}-\gamma 
   \cosh \left(2 \Delta(t)\right)\right)
   }{3 \left(\gamma ^2-2\right) \cosh
   \left(2 \Delta(t)\right)-2
   \sqrt{6} \gamma }\rightarrow \frac{2 \gamma ^2}{2-\gamma
   ^2},\\
z=\frac{\sqrt{6} \sqrt{6-\gamma ^2}
   \sinh \left(2 \Delta(t)\right)}{2
   \sqrt{6} \gamma -3 \left(\gamma
   ^2-2\right) \cosh \left(2 \Delta(t)\right)}\rightarrow \frac{\sqrt{4-\frac{2 \gamma
   ^2}{3}}}{2-\gamma ^2}.
\end{align} 
\end{subequations}
Thus, the point $J_7$ is approached by the solution \eqref{cosmol_sol} as the time  goes forward. 

To end this section we substitute the values of $y,z$ at $J_7$ in the equation \eqref{exact}. It follows by continuity that the solutions near $J_7$ satisfies the rates
\begin{subequations}
\label{asymptotic_J7}
\begin{align}
&\sigma (t)|_{J_7}= \sigma_0 e^{\frac{\gamma ^2 t
   \sqrt{U_0}}{\sqrt{2}
   \sqrt{6-\gamma ^2}}},\\
& a(t)|_{J_7}= a_0 e^{\frac{\left(2-\gamma
   ^2\right) t \sqrt{U_0}}{\sqrt{2}
   \sqrt{6-\gamma ^2}}},\\
&H(t)|_{J_7}= \frac{\left(2-\gamma ^2\right)
   \sqrt{U_0}}{\sqrt{2}
   \sqrt{6-\gamma ^2}}.
	\end{align}
\end{subequations}
	Now, comparing term by term the equations \eqref{asymptotic_J7} and \eqref{solutions-induced-gravity_2}, and identifying $\sigma_0=e^{\sqrt{\frac{3}{2}} \gamma 
   \left(c_2-c_3\right)}, a_0=e^{\sqrt{\frac{3}{2}} \gamma  \left(c_3-c_2\right)+c_2+c_3}$, $c_1= -\frac{\ln \left(4^{\frac{\gamma
   ^2}{6-\gamma ^2}}\right)
   \sqrt{\left(6-\gamma ^2\right)
   U_0}}{12 \sqrt{2} \gamma ^2
   U_0}$,
	it follows 
\begin{align}
&\frac{a(t)-a(t)|_{J_7}}{a(t)}=1-e^{-\frac{\left(\gamma ^2-2\right) t
   \sqrt{U_0}}{\sqrt{2}
   \sqrt{6-\gamma ^2}}} \sinh
   ^{-\frac{\sqrt{6} \gamma +2}{\sqrt{6}
   \gamma +6}}\left(\Delta(t)\right) \cosh
   ^{-\frac{\sqrt{6} \gamma -2}{\sqrt{6}
   \gamma -6}}\left(\Delta(t)\right)\rightarrow 0,\\
	&\frac{\sigma(t)-\sigma(t)|_{J_7}}{\sigma(t)}=1-e^{\frac{\gamma ^2 t
   \sqrt{U_0}}{\sqrt{2}
   \sqrt{6-\gamma ^2}}} \sinh
   ^{\frac{\sqrt{6} \gamma }{\sqrt{6}
   \gamma +6}}\left(\Delta(t)\right) \cosh
   ^{\frac{\sqrt{6} \gamma }{\sqrt{6}
   \gamma -6}}\left(\Delta(t)\right)\rightarrow 0,\\
&H(t)-H(t)|_{J_7}=\frac{\sqrt{U_0} \left(3 \gamma
   ^2-3 \left(\gamma ^2-2\right) \coth
   \left(\Delta(t)\right)+2
   \sqrt{6} \gamma 
   \text{csch}\left(\Delta(t)\right)-6\right
   )}{3 \sqrt{2} \sqrt{6-\gamma ^2}}\rightarrow 0.
\end{align}
Thus, we can use the asymptotics \eqref{asymptotic_J7} as approximations for the solutions near $J_7$. 
Summarizing, all the asymptotic results are consistent. 

In this example the ``intermediate accelerated'' solution does not exist, and the attractor
solution has an asymptotic de Sitter-like evolution law for the scale factor. 

\section{Field Equations in the Einstein's frame}\label{SECT:3}

Under the conformal transformation \cite{DabrowskiGareckiBlaschke2009}:
\begin{align}
\label{conformal}
\mathfrak{g}_{\mu\nu}= \Phi  {g}_{\mu\nu}\ \ \ \ \ \ \textrm{where}\ \ \ \ \ \ \Phi= e^{\gamma\chi},
\end{align} we can express the Jordan action \eqref{Jordan_action}  in the Einstein frame as:
\begin{align}
\label{action_1}
S_{EF}=\int\sqrt{-\mathfrak{g}}\left(\frac{\mathfrak{R}}{2}-\frac{1}{2}\mathfrak{g}^{\mu\nu}\partial_{\mu}\chi\partial_{\nu}\chi-\mathfrak{U}(\chi)-\frac{1}{2}e^{-\gamma \chi} \mathfrak{g}	^{\mu\nu}\partial_{\mu}\phi\partial_{\nu}\phi-e^{-2\gamma\chi}V(\phi)\right)d^4x,
\end{align}
where $\mathfrak{U}(\chi)=e^{-2 \gamma \chi} U(e^{\gamma \chi} )=U_0 e^{-\lambda_U \chi}$. In this frame $\chi$ is interpreted as a  conventional scalar field and $\phi$  is now coupled with $\chi$. For a discussion of the  equivalence between the two frames, see for example \cite{MagnanoFerrarisFrancaviglia1990, MagnanoSokolowski1994, Cotsakis1993, Teyssandier1995, Schmidt1995, Cotsakis1995, CapozzielloRitisMarino1997, FaraoniGunzigNardone1999, Faraoni2007, FaraoniNadeau2007,RomeroFonseca-NetoPucheu2012, QuirosGarcia-SalcedoAguilarEtAl2013}  and references therein. 

It is worth mentioning that in order for the conformal transformation to the Einstein's frame be well defined it is required that the scalar field $\Phi$ satisfies $\Phi>0$. However, it can asymptotically evolves to its minimum value $\Phi=0$. This implies that $\chi\rightarrow-\infty$ asymptotically,
and also, as we will see in the next section, the dynamical system variable $\epsilon$ tends asymptotically to zero at the fixed points.
	
By considering a flat Friedmann-Lema\^{\i}tre-Robertson-Walker (FLRW) metric
\begin{equation}\label{FRW_1}
d \hat{s}^2=-d\mathfrak{t}^2 + \mathfrak{a}(\mathfrak{t})^2\left[dr^2 +r^2 \left(d\theta^2 +\sin^2\theta d\varphi^2\right) \right],
\end{equation}
the field equations become:
\begin{subequations}
	\label{EQS_1}
\begin{align}
&\ddot{\chi}+3\mathcal{H}\dot{\chi}+\frac{\gamma}{2}e^{-\gamma \chi}\dot{\phi}^2-2\gamma e^{-2\gamma\chi}V(\phi)+\mathfrak{U}'(\chi)=0,\\
&\ddot{\phi}+3\mathcal{H}\dot{\phi}-\gamma\dot{\phi}\dot{\chi}+e^{-\gamma\chi}V'(\phi)=0,\\
&3\mathcal{H}^2- \left(\frac{1}{2}\dot{\chi}^2+\frac{1}{2}e^{-\gamma\chi}\dot{\phi}^2+e^{-2\gamma\chi}V(\phi)+\mathfrak{U}(\chi)\right)=0,\label{Fried_EF}\\
&2\dot {\mathcal{H}}+3 \mathcal{H}^ 2+ \left(\frac{1}{2}\dot{\chi}^2+\frac{1}{2}e^{-\gamma\chi}\dot{\phi}^2-\mathfrak{U}(\chi)-e^{-2\gamma\chi	}V(\phi)\right)=0,
\end{align}
\end{subequations}
where now the dot means derivative with respect the conformal time $\mathfrak{t}$. 

The relations with the quantities in the Jordan frame is through:
	\begin{align*}
	&\frac{d \mathfrak{t}}{d t}= \sqrt{\Phi},\quad \mathfrak{a}=\sqrt{\Phi} a,\quad \mathcal{H}=\frac{\dot{\mathfrak{a}}}{\mathfrak{a}}\equiv \frac{1}{\sqrt{\Phi}}\left[H+\frac{d}{dt}\left(\ln \sqrt{\Phi}\right)\right]=\frac{1}{\sqrt{\Phi}}\left[H+\frac{1}{2 \Phi}\frac{d\Phi}{dt}\right].
	\end{align*}

We define the effective energy densities $\mu_1, \mu_2$, the effective pressures  $\mathfrak{p}_1$ and $\mathfrak{p}_2$, and the coupling term $Q$ as: \cite{FaraoniDentSaridakis2014}:
\begin{subequations}
	\begin{align}
	&\mu_1=\frac{1}{2}{\dot{\chi}}^2+U(\chi),\\
	&\mu_2=\frac{1}{2}e^{-\gamma \chi}{\dot{\phi}}^2+e^{-2\gamma \chi} V(\phi),\\
	&\mathfrak{p}_1=\frac{1}{2}{\dot{\chi}}^2-U(\chi),\\
	&\mathfrak{p}_2=\frac{1}{2}e^{-\gamma \chi}{\dot{\phi}}^2-e^{-2\gamma \chi} V(\phi),\\
	&Q=\frac{1}{2}\gamma e^{-2\gamma\chi}\dot{\chi}\left[4 V(\phi)-e^{\gamma\chi}{\dot{\phi}}^2\right].
	\end{align}
	\end{subequations}
Then, the system \eqref{EQS_1} is equivalent to 
\begin{subequations}
	\begin{align}
	& \dot{\mu}_1+3\mathcal{H}(\mu_1+\mathfrak{p}_1)=Q,\\
	& \dot{\mu}_2+3\mathcal{H}(\mu_2+\mathfrak{p}_2)=-Q,\\
	& \mathcal{H}^2=\frac{1}{3}\left(\mu_1+\mu_2\right),\\
	&\dot{\mathcal{H}}=-\frac{1}{2}\left(\mu_1+\mathfrak{p}_1+\mu_2+\mathfrak{p}_2\right),
	\end{align}
\end{subequations}
which is interpreted as the Einstein equations for two scalar fields coupled in a non-standard way.

\subsection{Dynamical system analysis}\label{dyn_E}

In this section we present the above system by defining proper dynamical variables. Next, we examine the stability and discuss the properties of the solutions in this frame.

\subsubsection{Finite analysis}

Let us first note that the Friedmann equation \eqref{Fried_EF} can be expressed as:
\begin{equation}\frac{e^{-\gamma  \chi} {\dot\phi}^2}{6 \mathcal{H}^2}+\frac{\mathfrak{U}(\chi)}{3 \mathcal{H}^2}+\frac{{\dot\chi
   }^2}{6 \mathcal{H}^2}+\frac{e^{-2 \gamma  \chi} V(\phi)}{3 \mathcal{H}^2}=1.
	\end{equation}

This expression suggests the introduction of the following variables
\begin{equation}\label{vars_EF}
\epsilon=e^{\frac{\gamma}{2}\chi},\; x_1=\frac{e^{-\frac{\gamma}{2}\chi}\dot\phi}{\sqrt{6}\mathcal{H}},\;  y_1=\frac{\dot\chi}{\sqrt{6}\mathcal{H}},\; z_1=\frac{\sqrt{\mathfrak{U}(\chi)}}{\sqrt{3}\mathcal{H}}.
\end{equation}
The Friedmann equation \eqref{Fried_EF} leads to 
\begin{equation}
	\label{FRIED2}
\frac{V(\phi)}{3 \mathcal{H}^2\epsilon^4}+ x_1^2+y_1^2+z_1^2=1. 
\end{equation} 
The evolution equations for the variables \eqref{vars_EF} are given by: 
\begin{subequations}
	\label{DS_Einstein}
\begin{align}
& x_1'=\frac{1}{2} x_1 \left(6 x_1^2+\sqrt{6} \gamma  y_1+6
y_1^2-6\right)-\sqrt{\frac{3}{2}}  \lambda_V \epsilon 
\left(x_1^2+y_1^2+z_1^2-1\right),\\
& y_1'=\sqrt{\frac{3}{2}} \left(2 \gamma -3 \gamma 
x_1^2+\sqrt{6} \left(x_1^2-1\right) y_1-2 \gamma  y_1^2+\sqrt{6}
y_1^3+z_1^2 ( \lambda_U-2 \gamma )\right),\\
& z_1'=\frac{1}{2} z_1 \left(6
x_1^2+y_1 \left(6 y_1-\sqrt{6}  \lambda_U\right)\right),\\
& \epsilon'= \sqrt{\frac{3}{2}} \gamma  y_1 \epsilon,
\end{align}
\end{subequations}
which are defined on the phase space 
\begin{align}
\Xi:=\{(x_1,y_1,z_1,\epsilon)\in\mathbb{R}^4: x_1^2+y_1^2+z_1^2\leq 1, z_1\geq 0\},
\end{align}
where the comma denotes derivative with respect $\hat{\tau}=\ln \mathfrak{a}$.

The relation with the variables \eqref{vars} is given by 
\begin{subequations}\label{relation_vars}
	\begin{align}
	 &\epsilon=\epsilon, \quad x_1=\frac{2 x}{2+y}, \quad
    y_1=\frac{\sqrt{\frac{2}{3}}y}{(2+y) \gamma}, \quad z_1=\frac{2 z}{2+y},
\text{with inverse}\\
    &\epsilon=\epsilon,\quad x=\frac{2 x_1}{2-\sqrt{6} \gamma  y_1}, \quad
    y=\frac{6 \gamma  y_1}{\sqrt{6}-3 \gamma  y_1}, \quad
    z=\frac{2 z_1}{2-\sqrt{6} \gamma  y_1}.
	\end{align}
	\end{subequations}
The conditions $y\neq -2$ and $y_1\neq  \sqrt{\frac{2}{3 \gamma^2}}$ guarantee that the direct transformation  and its inverse given by Eqs. \eqref{relation_vars} are non singular.

The relation between the coordinates defined in Einstein frame by \eqref{vars_EF}  and the Poincar\'e coordinates given by \eqref{POINCAREJBD} is:
\begin{subequations}
	\label{POINCARE_EF}
	\begin{align}
	&X=-\frac{2 x_1}{\left(\sqrt{6} \gamma  y_1-2\right) \sqrt{\frac{2 \left(x_1^2+4
				\sqrt{6} \gamma  y_1+z_1^2-4\right)}{\gamma  y_1 \left(3 \gamma  y_1-2
				\sqrt{6}\right)+2}+5}},\\
	&Y=\frac{6 \gamma  y_1}{\left(\sqrt{6}-3 \gamma  y_1\right)
		\sqrt{\frac{2 \left(x_1^2+4 \sqrt{6} \gamma  y_1+z_1^2-4\right)}{\gamma  y_1 \left(3
				\gamma  y_1-2 \sqrt{6}\right)+2}+5}},\\
	&Z=-\frac{2 z_1}{\left(\sqrt{6} \gamma 
		y_1-2\right) \sqrt{\frac{2 \left(x_1^2+4 \sqrt{6} \gamma  y_1+z_1^2-4\right)}{\gamma
				y_1 \left(3 \gamma  y_1-2 \sqrt{6}\right)+2}+5}}.
	\end{align}
\end{subequations}

The cosmological parameters are given by:
\begin{subequations}
	\label{cosmolparEF}
	\begin{align}
		&\hat{\Omega}_1\equiv \frac{{\mu}_1}{3\mathcal{\mathcal{H}}^2}= y_1^2+z_1^2,\\
		&\hat{\Omega}_2\equiv \frac{{\mu}_2}{3\mathcal{\mathcal{H}}^2}=1-y_1^2-z_1^2,\\
		& \hat{q}\equiv -1-\frac{\dot{\mathcal{H}}}{\mathcal{H}^2}=-1+3(x_1^2+y_1^2),\\
		& \hat{w}_{\text{tot}}\equiv -1-\frac{2 \dot{\mathcal{H} }}{3 \mathcal{H}^2}= -1+2(x_1^2+y_1^2).
	\end{align}
\end{subequations}

\paragraph{Fixed points and stability in the Einstein frame.}\label{A2}

				The critical points, eigenvalues and stability conditions for the system in the Einstein frame \eqref{DS_Einstein} are:
				\begin{enumerate}
					\item $E_{1,2}: (x_1=0, y_1=\pm 1, z_1=0, \epsilon=0)$. They always exist. The eigenvalues are: $$\left\{\pm\sqrt{\frac{3}{2}} \gamma ,\pm \sqrt{\frac{3}{2}} \gamma ,6 \mp 2 \sqrt{6} \gamma
					,3 \mp \sqrt{\frac{3}{2}}\lambda_U\right\}.$$  $E_2$ is a source for $\lambda_U<\sqrt{6}, 0<\gamma <\sqrt{\frac{3}{2}}$, or a saddle otherwise. $E_1$ is always a saddle. These points are related with $J_{1,2}$ through \eqref{relation_vars}. 
					\item $E_3: \left(0,\sqrt{\frac{2}{3}} \gamma ,0,0\right)$. It exists for $0<\gamma^2\leq \frac{3}{2}$. The eigenvalues are: $$\left\{\gamma ^2,2 \gamma ^2-3,3 \left(\gamma ^2-1\right),\gamma  (2 \gamma -\lambda_U)\right\}.$$ Whenever it exists it is always a saddle. This point is related to $J_3$ through \eqref{relation_vars}. 
					\item $E_4: \left(0,0,\frac{\sqrt{2} \sqrt{\gamma }}{\sqrt{2 \gamma -\lambda_U}},0\right)$. It exists for $\gamma>0,\lambda_U\leq 0$. The eigenvalues are $$\left\{0,-3,\frac{1}{2} \left(-3-\sqrt{24 \gamma  {\lambda_U}+9}\right),\frac{1}{2}
					\left(-3+\sqrt{24 \gamma  {\lambda_U}+9}\right)\right\}.$$ It is nonhyperbolic with a 3D stable manifold for $\lambda_U<0, \gamma>0.$ Thus, it has a large probability to attract the universe at late times. The full stability analysis requires the application of the center manifold theorem (the analysis is done in subsection \ref{CENTER_H4_bd}). This point is analogous to $J_4$ studied before. 
					\item $E_{5,6}: \left(\pm \sqrt{\frac{2}{3}},0,0,\pm \frac{2}{\lambda_U}\right)$. They always exist. 
					The eigenvalues are $$\left\{-1,2,\frac{1}{2} \left(-\sqrt{1-24 \gamma ^2}-1\right),\frac{1}{2} \left(\sqrt{1-24
						\gamma ^2}-1\right)\right\}.$$ Thus, they are always saddle. These points are completely analogous to $J_{5,6}$  examined before. 
					\item $E_7: \left(0,\frac{{\lambda_U}}{\sqrt{6}},\sqrt{1-\frac{{\lambda_U}^2}{6}},0\right)$. It exists for $-\sqrt{6}\leq {\lambda_U}\leq \sqrt{6}$.  The eigenvalues are $$\left\{\frac{\gamma  {\lambda_U}}{2},\frac{1}{2} ({\lambda_U} (\gamma
					+{\lambda_U})-6),\frac{1}{2} \left({\lambda_U}^2-6\right),{\lambda_U}
					({\lambda_U}-2 \gamma )\right\}.$$ Thus, it is a saddle. This point is related to $J_7$ through \eqref{relation_vars}.
					\item $E_{8,9}: \left(\pm\frac{\sqrt{{\lambda_U} (\gamma +{\lambda_U})-6}}{\gamma +\lambda_U},\frac{\sqrt{6}}{\gamma +{\lambda_U}},\frac{\sqrt{\gamma }}{\sqrt{\gamma
							+{\lambda_U}}},0\right)$. They exist for $0 \leq \lambda_U\leq \sqrt{6}, \gamma>\frac{6-\lambda_U^2}{\lambda_U}$ or $\gamma \geq 0, \lambda_U>\sqrt{6}$. The eigenvalues are $$\left\{\frac{3 \gamma }{\gamma +{\lambda_U}},\frac{\lambda_1}{2 (\gamma
						+{\lambda_U})^{7/2}},\frac{\lambda_2}{2 (\gamma +\lambda_U)^{7/2}},\frac{\lambda_3}{2 (\gamma +{\lambda_U})^{7/2}}\right\},$$ where $\lambda_{1,2,3}$ are the roots of: 
					$P(\lambda)=12 \gamma  \lambda  \left(\left(\gamma ^2-12\right) {\lambda_U}+2 \gamma  \lambda_U^2+6 \gamma +{\lambda_U}^3\right) (\gamma +{\lambda_U})^{11/2}+\lambda ^3
					\sqrt{\gamma +{\lambda_U}}+6 \lambda ^2 (5 \gamma -2 {\lambda_U}) (\gamma
					+{\lambda_U})^3+144 \gamma  (2 \gamma -{\lambda_U}) ({\lambda_U} (\gamma
					+{\lambda_U})-6) (\gamma +{\lambda_U})^9$. These points are related to $J_{8,9}$ through \eqref{relation_vars}. Observe that from the existence conditions it follows that the first eigenvalue is always positive, so, these critical points are either saddles or sources, but never an attractor.

					\item $E_{10,11}: \left(\pm \frac{\sqrt{\frac{2}{3}} \sqrt{\gamma ^2-1}}{\gamma },\frac{\sqrt{\frac{2}{3}}}{\gamma
					},0,0\right)$. They exist for $-\frac{3}{2}<\omega_0<-\frac{1}{2}$. The eigenvalues are $$\left\{1,-\frac{1}{2}\left(1-\sqrt{25-24\gamma^2}\right),-\frac{1}{2}\left(1+\sqrt{25-24\gamma^2}\right), 2-\frac{\lambda_U}{\gamma} \right\}.$$ They are always saddles. Since the transformation \eqref{vars} is not well defined for $y_1=\frac{\sqrt{\frac{2}{3}}}{\gamma}$, there are no equivalent points in the Jordan frame at the finite region of the phase space (see the subsection  \ref{A3}). However, using the relation \eqref{POINCARE_EF} we get that $E_{10,11}$ maps onto $Q_{3,4}$ which are located in the region at infinity (see Table \ref{Tabinfinity}). 
				\end{enumerate}
				
The existence and stability conditions for the critical points of 
\eqref{DS_Einstein} are displayed in Table \ref{Tab1}. The relevant cosmological parameters and the description of the critical points in terms of their stability are shown in Table \ref{Tab1b} \footnote{As in the Jordan frame, the leading terms of the Hubble parameter in the neighborhood of the critical points are also shown.}.
				\begin{landscape}{\begin{table*}
			\begin{tabular}{@{ }c@{ }c@{ }c@{ }c@{ }}
				\hline \rule[-2ex]{0pt}{5.5ex} Label & $x_1, y_1, z_1, \epsilon$ & Existence & Stability \\ 
				\hline \rule[-2ex]{0pt}{5.5ex} $E_{1,2}$  & $\left(0, \pm 1, 0, 0\right)$ & always &  $E_1$ is  a saddle   \\ 
				& &  & $E_2$ is a source for $\lambda_U<\sqrt{6}, \omega_0>-\frac{5}{6}$, \\ 
				& & & saddle otherwise     \\
				\hline \rule[-2ex]{0pt}{5.5ex} $E_3$  & $\left(0,\sqrt{\frac{2}{3}} \gamma ,0,0\right)$ & $\omega_0>-\frac{5}{6}$  & saddle    \\ 
				
				\hline \rule[-2ex]{0pt}{5.5ex} $E_4$ & $\left(0,0,\frac{\sqrt{2} \sqrt{\gamma }}{\sqrt{2 \gamma -\lambda_U}},0\right)$ & $\omega_0>-\frac{3}{2},\lambda_U\leq 0$ & sink for  $\gamma>0, \lambda_V\neq 0,  \lambda_U<0$. \\ 
				\hline \rule[-2ex]{0pt}{5.5ex} $E_{5,6}$ & $\left(\pm \sqrt{\frac{2}{3}},0,0,\pm \frac{2}{\lambda_U}\right)$ & always  & saddle   \\ 
				\hline \rule[-2ex]{0pt}{5.5ex}  $E_7$ & $\left(0,\frac{{\lambda_U}}{\sqrt{6}},\sqrt{1-\frac{{\lambda_U}^2}{6}},0\right)$ & $-\sqrt{6}\leq {\lambda_U}\leq \sqrt{6}$ & saddle   \\ 
				\hline \rule[-2ex]{0pt}{5.5ex}  $E_{8,9}$ & $\left(\pm\frac{\sqrt{{\lambda_U} (\gamma +{\lambda_U})-6}}{\gamma +\lambda_U},\frac{\sqrt{6}}{\gamma +{\lambda_U}},\frac{\sqrt{\gamma }}{\sqrt{\gamma
						+{\lambda_U}}},0\right)$ &   $\omega_0>-\frac{3}{2}, \sqrt{\frac{12 \omega_0+\sqrt{24 \omega_0+37}+19}{2\omega_0+3}}\leq \lambda_U\leq \sqrt{6}$ or &   \\ 
				& & $\omega_0\geq -\frac{3}{2}, \lambda_U>\sqrt{6}$ & numerical inspection    \\ 
				\hline \rule[-2ex]{0pt}{5.5ex} $E_{10,11}$ &$\left(\mp \frac{\sqrt{\frac{2}{3}} \sqrt{\gamma ^2-1}}{\gamma },\frac{\sqrt{\frac{2}{3}}}{\gamma
				},0,0\right)$  & $-\frac{3}{2}<\omega_0<-\frac{1}{2}$ & saddle \\
				\hline 
			\end{tabular} 
			\caption{\label{Tab1} The existence and stability conditions for the critical points of 
				\eqref{DS_Einstein}.}
		\end{table*}}
	\end{landscape}
	
	\begin{landscape}
	\begin{table*}
	\label{EFObs1x} 
		\begin{tabular}{|@{ }c@{ }|c@{ }|c@{ }|c@{ }|c@{ }|c@{ }|c@{ }|}
			\hline \rule[-2ex]{0pt}{5.5ex} Label & $\hat{\Omega}_1$ & $\hat{\Omega}_2$ & $\hat{q}$ & $\hat{w}_{\text{tot}}$ & Description & $\mathcal{H}(\mathfrak{t})$  \\ 
			\hline \rule[-2ex]{0pt}{5.5ex} $E_{1,2}$   & $1$& $0$ & $2$   & $1$ & stiff matter & $\frac{\mathcal{H}_0}{1+3 \mathcal{H}_0 (\mathfrak{t}-\mathfrak{t}_0)}.$ \\ 
			\hline \rule[-2ex]{0pt}{5.5ex} $E_3$  & $\frac{2 \gamma ^2}{3}$ & $1-\frac{2 \gamma ^2}{3}$ & $2 \gamma ^2-1$ & $\frac{4 \gamma ^2}{3}-1$ & scaling solution. & $\frac{\mathcal{H}_0}{1+2\gamma^2 \mathcal{H}_0 (\mathfrak{t}-\mathfrak{t}_0)}.$\\   
			\hline \rule[-2ex]{0pt}{5.5ex} $E_4$ & $\frac{2 \gamma }{2 \gamma -{\lambda_U}}$ & $-\frac{{\lambda_U}}{2 \gamma -{\lambda_U}}$ & $-1$ & $-1$ & Intermediate accelerated &  $\simeq \alpha_2 p_2 \mathfrak{t}^{p_2-1}.$ \\
			&  &  &  &  & $\mathfrak{a}(\mathfrak{t})\simeq e^{\alpha_2  \mathfrak{t}^{p_2}}$, $\alpha_2>0, 0<p_2<1.$&\\ 
			\hline \rule[-2ex]{0pt}{5.5ex} $E_{5,6}$ & $0$ & $1$  & $1$ & $\frac{1}{3}$ & radiation-like. & $\frac{\mathcal{H}_0}{1+2\mathcal{H}_0 (\mathfrak{t}-\mathfrak{t}_0)}.$\\ 
			\hline \rule[-2ex]{0pt}{5.5ex}  $E_7$ & $1$ & $0$ & $\frac{1}{2} \left(\lambda_U^2-2\right)$ & $\frac{1}{3} \left(\lambda_U^2-3\right)$ & quintessence-dominated. & $\frac{2\mathcal{H}_0}{2+\lambda_U^2\mathcal{H}_0 (\mathfrak{t}-\mathfrak{t}_0)}.$ \\ 
			\hline \rule[-2ex]{0pt}{5.5ex}  $E_{8,9}$ & $\frac{\gamma  (\gamma +{\lambda_U})+6}{(\gamma +{\lambda_U})^2}$ & $\frac{{\lambda_U} (\gamma +{\lambda_U})-6}{(\gamma
				+{\lambda_U})^2}$ & $2-\frac{3 \gamma }{\gamma +{\lambda_U}}$ & $1-\frac{2 \gamma }{\gamma +{\lambda_U}}$ & scaling solution. & $\frac{\mathcal{H}_0}{1+\frac{3\mathcal{H}_0 \lambda_U}{\gamma+\lambda_U}(\mathfrak{t}-\mathfrak{t}_0)}.$  \\  
			\hline \rule[-2ex]{0pt}{5.5ex} $E_{10,11}$ & $\frac{2}{3 \gamma ^2}$ & $1-\frac{2}{3 \gamma ^2}$ & $1$ & $\frac{1}{3}$ & radiation-like.  & $\frac{\mathcal{H}_0}{1+2\mathcal{H}_0 (\mathfrak{t}-\mathfrak{t}_0)}.$ \\\hline 
		\end{tabular} 
		\caption{\label{Tab1b} Description of the cosmological parameters \eqref{cosmolparEF} of the critical points of 
			\eqref{DS_Einstein}.}
	\end{table*}
	\end{landscape}

\paragraph{Center manifold analysis for the intermediate accelerated solution $E_4$}\label{CENTER_H4_bd}

				In order to investigate the stability of the center manifold for $E_4$ we introduce the new variables
				\begin{subequations}
					\label{centerB}
					\begin{align}
					&u=\epsilon,\\
					&v_1=\frac{\lambda_U \lambda_V \epsilon }{\sqrt{6} (2 \gamma
						-\lambda_U)}+x_1,\\
					&v_2=\frac{\sqrt{\gamma } \left(\sqrt{3}-\sqrt{8 \gamma 
							\lambda_U+3}\right)}{\sqrt{4 \gamma -2\lambda_U} \sqrt{8 \gamma 
							\lambda_U+3}}+\frac{\sqrt{\gamma }\lambda_U y_1}{\sqrt{2 \gamma
							-\lambda_U} \sqrt{8 \gamma \lambda_U+3}}-\frac{z_1
						\left(\sqrt{3}-\sqrt{8 \gamma \lambda_U+3}\right)}{2 \sqrt{8 \gamma \lambda_U+3}},\\
					&v_3=-\frac{\sqrt{\gamma } \left(\sqrt{8 \gamma \lambda_U+3}+\sqrt{3}\right)}{\sqrt{4 \gamma -2\lambda_U} \sqrt{8 \gamma  
							\lambda_U+3}}-\frac{\sqrt{\gamma }\lambda_U y_1}{\sqrt{2 \gamma -\lambda_U}
						\sqrt{8 \gamma \lambda_U+3}}+\frac{z_1 \left(\sqrt{8 \gamma \lambda_U+3}+\sqrt{3}\right)}{2 \sqrt{8 \gamma \lambda_U+3}},
					\end{align}
				\end{subequations}
				to obtain
				{{\small
						\begin{align}
						\left(
						\begin{array}{c}
						u' \\
						v_1' \\
						v_2' \\
						v_3'
						\end{array}
						\right)= \left(
						\begin{array}{cccc}
						0 & 0 & 0 & 0 \\
						0 & -3 & 0 & 0 \\
						0 & 0 & -\frac{1}{2} \left(3+\sqrt{24 \gamma  \lambda_{U}+9}\right) & 0 \\
						0 & 0 & 0 & -\frac{1}{2} \left(3-\sqrt{24 \gamma  \lambda_{U}+9}\right) \\
						\end{array}
						\right) \left(
						\begin{array}{c}
						u \\
						v_1 \\
						v_2 \\
						v_3
						\end{array}
						\right) + \left(
						\begin{array}{c}
						f \\
						g_1 \\
						g_2 \\
						g_3
						\end{array}
						\right)
						\end{align}
					}} 
					where $(f,g_1,g_2, g_3)^T$ is a vector of higher order terms. 
					
					Since the center subspace of the origin is tangent to the $\epsilon$-axis, it follows that the center manifold of the origin is given locally by the graph
					\begin{align}
					&\Big\{(u, v_1,v_2,v_3): v_1=h_1(u), v_2=h_2(u), v_3=h_3(u), \nonumber \\ &  h_1(0)=h_2(0)=h_3(0)=0, \nonumber \\ & h_1'(0)=h_2'(0)=h_3'(0)=0, |u|<\delta \Big\},
					\end{align}
					where $\delta$ is a positive small enough constant. 
					The functions $h_i, i=1,2,3$ satisfy a set of quasilinear ordinary differential equations which can be expressed symbolically as
					\begin{equation}
					\label{h'sB}
					\Big[h_i'(u)u'-v_i'\Big]\Big|_{v_i=h_i(u)}=0, i=1,2,3,
					\end{equation}
					where one must substitute $u', v_1', v_2', v_3'$ through \eqref{centerB} and use the replacement $v_1\rightarrow h_1(u), v_2\rightarrow h_2(u)$
					and $v_3\rightarrow h_3(u)$.
					
					Setting
					\begin{subequations}
						\begin{align}
						& h_1(u)=a_{11} u^2 + a_{12} u^3 +\mathcal{O}\left(u\right)^4, \\ & h_2(u)=a_{2 1} u^2 + a_{2 2} u^3 +\mathcal{O}\left(u\right)^4, \\ & h_3(u)=a_{3 1} u^2 + a_{3 2} u^3 +\mathcal{O}\left(u\right)^4, 
						\end{align}
					\end{subequations}
					in \eqref{h'sB}, equating to zero all the coefficients of equal powers of $u$, and solving for the $a_{i j}$'s we get up to fourth order:  
					
					\begin{subequations}
						\begin{align}
						& a_{1 1}=0 , a_{1 2}=-\frac{\lambda_U \lambda_V^3 (\gamma  \lambda_U+6)}{6 \sqrt{6} (\lambda_U-2 \gamma )^3} ,\\
						&    a_{2 1}=-\frac{\sqrt{\gamma }
							\lambda_U^2 \lambda_V^2 \left(\sqrt{3} \gamma  \lambda_U-\sqrt{8
								\gamma  \lambda_U+3}+\sqrt{3}\right)}{\sqrt{2} (2 \gamma -\lambda_U)^{5/2}
							\left(8 \sqrt{3} \gamma  \lambda_U+3 \left(\sqrt{8 \gamma  \lambda_U+3}+\sqrt{3}\right)\right)} ,a_{2 2}=0,\\
						& a_{3 1}=-\frac{\sqrt{\gamma } \lambda_U^2
							\lambda_V^2 \left(\sqrt{3} \gamma  \lambda_U+\sqrt{8 \gamma  \lambda_U+3}+\sqrt{3}\right)}{\sqrt{2} (2 \gamma -\lambda_U)^{5/2} \left(8 \sqrt{3} \gamma
							\lambda_U-3 \sqrt{8 \gamma  \lambda_U+3}+3 \sqrt{3}\right)},
						a_{3 2}=0
						.
						\end{align}
					\end{subequations}
					Henceforth, although the definition of the center manifold is quite different as for $J_4$, the dynamics on the center manifold is given by the same equation \eqref{CENTERA}, i.e.,   
					\begin{equation}
					u'=\frac{\gamma  \lambda_U \lambda_V^2 u^3}{2 (\lambda_U-2 \gamma
						)^2}+\mathcal{O}\left(u\right)^5.
					\end{equation}
					Neglecting the fifth-order terms and integrating we find that 
					\begin{equation}
					\label{u_expansion_EFrame}
					u(\hat{\tau})=\pm \frac{\sqrt{-(\lambda_U-2 \gamma )^2}}{\sqrt{2 c_1 (\lambda_U-2 \gamma
							)^2+\gamma  \lambda_U \lambda_V^2\hat{\tau} }},
					\end{equation}
					where $c_1$ is an integration constant that must be negative in order for $u$ to be real-valued. 
					Thus, as before,  for $\gamma>0, \lambda_U\notin\{0, 2\gamma\}, \lambda_V\neq 0$,  it follows that the origin, and then $E_4$, is stable provided $\lambda_U<0$.

\paragraph{Special case: $\lambda_U=0$, $\mathfrak{U}(\chi)= U_0$.}
					
					For  $\lambda_U=0$, i.e., $\mathfrak{U}(\chi)= U_0$ we introduce the new variables
					\begin{subequations}
						\begin{align}
						& u_1=\epsilon, \\
						& u_2=z_1-1,\\
						& v_1= x_1,\\
						& v_2=y_1+2 \sqrt{\frac{2}{3}} \gamma  (z_1-1). 
						\end{align}
					\end{subequations}
					{{\small
							\begin{align}
							\label{centerBspecial}
							\left(
							\begin{array}{c}
							u_1' \\
							u_2' \\
							v_1' \\
							v_2'
							\end{array}
							\right)= \left(
							\begin{array}{cccc}
							0 & 0 & 0 & 0 \\
							0 & 0 & 0 & 0 \\
							0 & 0 & -3 & 0 \\
							0 & 0 & 0 & -3 \\
							\end{array}
							\right) \left(
							\begin{array}{c}
							u_1 \\
							u_2 \\
							v_1 \\
							v_2
							\end{array}
							\right) + \text{higher order terms}.
							\end{align}
						}}

						Since the center subspace of the origin is tangent to the plane $u_1$-$u_2$, it follows that the center manifold of the origin is given locally by the graph
						\begin{align}
						&\Big\{(u_1, u_2, v_1,v_2): v_1=h_1(u_1,u_2), v_2=h_2(u_1,u_2), \nonumber \\ &  h_1(0,0)=h_2(0,0)=0, \mathbf{Dh}(0,0)=\mathbf{0}, u_1^2+u_2^2<\delta \Big\},
						\end{align}
						where $\mathbf{Dh}$ is the matrix of derivatives and $\delta$ is a positive small enough constant. 
						The functions $h_1, h_2$ satisfy a set of quasilinear partial differential equations which can be expressed symbolically as
						\begin{equation}
						\label{specialBh's}
						\Big[\frac{\partial{h_i(u)}}{\partial{u_1}}u_1'+\frac{\partial{h_i(u)}}{\partial{u_2}}u_2'-v_i'\Big]\Big|_{v_i=h_i(u)}=0, i=1,2,
						\end{equation}
						where one must substitute $u_1', u_2', v_1', v_2',$ through \eqref{centerBspecial} and use the replacement $v_1\rightarrow h_1(u_1,u_2), v_2\rightarrow h_2(u_1,u_2)$.
						
						Setting 
						\begin{subequations}
							\begin{align}
							& h_1=a_{11} u_1^2 + a_{12} u_1 u_2 +a_{22} u_2^2 +\mathcal{O}(3),\\
							& h_2=b_{11} u_1^2 + b_{12} u_1 u_2 +b_{22} u_2^2 +\mathcal{O}(3),
							\end{align}
						\end{subequations}
						and plugging back in  \eqref{specialBh's}, equating to zero all the coefficients of equal powers of $u_1$ and $u_2$, and solving for the $a_{i j}$'s and $b_{i j}$'s we get up to third order
						\begin{equation}
						\left(
						\begin{array}{ccc}
						a_{11} & a_{12} & a_{22} \\
						b_{11} & b_{12} & b_{22}\\
						\end{array}
						\right)=\left(
						\begin{array}{ccc}
						0 & -\sqrt{\frac{2}{3}} \lambda_V & 0 \\
						0 & 0 & \frac{1}{3} \sqrt{\frac{2}{3}} \gamma  \left(8 \gamma ^2-3\right) \\
						\end{array}
						\right).
						\end{equation}
						Thus, the dynamics on the center manifold is dictated by 
						\begin{subequations}
							\label{specialB_center}
							\begin{align}
							& u_1'=-2 \gamma ^2 u_1 u_2+\mathcal{O}(3),\\
							& u_2'=8 \gamma ^2 u_2^2 +\mathcal{O}(3),
							\end{align}
						\end{subequations}
						where $\mathcal{O}(3)$ denotes error terms of third order in the vector norm. 
						Neglecting the error terms and integrating out the system \eqref{specialB_center}
						we obtain
						\begin{equation}
						u_1=c_2
						\sqrt[4]{8 \gamma ^2 \tau +c_1}, u_2=-\frac{1}{8 \gamma ^2 \tau +c_1}.
						\end{equation}
						Finally, it follows that for $\lambda_U=0$, the origin, and then the point $E_4$ behaves as a saddle since the orbits departs from the origin along the $\epsilon$-direction as the time goes forward. 					
	
								\paragraph{Features of the critical points of the system \eqref{DS_Einstein}.}
								
Let us summarize the features of the critical points of the system \eqref{DS_Einstein} 
found in subsection \ref{A2}:

	\begin{enumerate}
		\item Solutions $E_{1,2}$ are stiff-like solutions dominated by the kinetic term of the quintessence field $\chi$. This points are saddle so, they are not late-time solutions.
		
		\item $E_3$ is a scaling solution where $\dot{\chi}^2$ has the same order of magnitude than the effective potential $V(\phi)/\epsilon^4$. 
		
		\item $E_4$ is a solution where the potential energy $\mathfrak{U}(\chi)$ of the quintessence scalar field and the effective potential $V(\phi)/\epsilon^4$  have the same order of magnitude. Besides $\hat{w}_{\text{tot}}=-1$, at the critical point. However, as we will prove later on subsection \ref{Eframe}, it does not represent generically a de Sitter solution. Indeed, the scale factor $\mathfrak{a}$ satisfy
		\begin{align}
		&\mathfrak{a}\simeq e^{\alpha_2 \mathfrak{t}^{p_2}}\; (\text{leading terms as } \mathfrak{t} \rightarrow \infty),
		\end{align} 
		where $\alpha_2>0$ and $0<p_2<1$. 
				
		\item $E_{5,6}$ is a saddle-like radiation dominated solution since $w_{\text{tot}}=\frac{1}{3}$. 
		\item $E_7$ represents the standard quintessence dominated solution. As a difference with the usual case \cite{CopelandLiddleWands1998}, it can not attract the universe at late times since it is a saddle. 
		\item $E_{8,9}$ represents a scaling solutions where the energy density of both scalars scales with the same order of magnitud. Neither one field, nor the other, dominates the dynamics. 
		\item  The critical points  $E_{10,11}$ mimics a radiation solution since $w_{\text{tot}}=\frac{1}{3}$. Using the relation \eqref{POINCARE_EF} we get that $E_{10,11}$ maps onto $Q_{3,4}$ (see Table \ref{Tabinfinity}). The existence conditions for these points lead to $\omega_0$-values which are lower than the observed values for the BD parameter according to the recent analysis \cite{Will2014, BertottiIessTortora2003, AcquavivaBaccigalupiLeachEtAl2005, NagataChibaSugiyama2004}.  
		 
	\end{enumerate}
	
	\subsubsection{Analysis at infinity}

Since the variable $\epsilon$ is unbounded we add to $x_1,y_1,z_1$ defined in \eqref{vars_EF} the new variable defined by $E=\frac{\epsilon}{1+\epsilon}$, and the new time variable $\check{T}$ given by  ${d \check{T}}\equiv (1-E)^{-1}{d\tau}$. 
The resulting system admits the set of non-hyperbolic equilibrium points \footnote{All the eigenvalues of the Jacobian matrix are zero.} at infinity $x_c^2+y_c^2+z_c^2=1$ and $E=1.$ That is the boundary of the phase space $\Xi$ but with $\Phi\rightarrow \infty$.

\subsection{Viability of the intermediate accelerated solution in the Einstein Frame }
 	\label{Eframe}
 	
 In the Einstein frame,  $E_4$ is probably the more interesting late-time solution. It represents a solution where the potential energy $\mathfrak{U}(\chi)$ of the quintessence scalar field and the effective potential $V(\phi)/\epsilon^4$  have the same order of magnitude. Besides $\hat{w}_{\text{tot}}=-1$ at the critical point. However, it does not represent generically a de Sitter solution in our scenario as we will prove in the following.
 	
 From the definition of $z_1$ it follows that 	at equilibrium point
 \begin{equation}
 	\label{approxH_EF}
 	\mathcal{H}=\frac{\sqrt{(2\gamma-\lambda_U)U_0}}{\sqrt{6\gamma}} e^{-\frac{\lambda_U}{2}\chi}.
 	\end{equation}
Since $\epsilon\equiv e^{\frac{\gamma}{2}\chi}$ tends to zero when $E_4$ is approached and $\gamma>0$, it follows that $\chi\rightarrow -\infty$ at late time. 
Now, under the conditions $\gamma>0, \lambda_V\neq 0, \lambda_U<0$, which imply the stability of $E_4$, it follows that $\mathcal{H}\rightarrow 0$ asymptotically.  Furthermore, since the deceleration parameter satisfy $\hat{q}\rightarrow -1$ at the critical point, it follows by continuity that $\dot{\mathcal{H}}\ll {\mathcal{H}}^2$ at late time, which means that $\dot{\mathcal{H}}$ tends to zero too. 
 	
Now, let us take advantage of the formula \eqref{u_expansion_EFrame}, which is valid up to fifth order, for obtaining some asymptotic expansions.
Since $\chi=\frac{1}{\gamma}\ln u^2$, then we result in
 	\begin{equation}
 	\label{approx1B}
 	\chi(\mathfrak{a})=\frac{1}{\gamma}\ln\left[\frac{(\lambda_U-2 \gamma )^2}{-2 c_1 (\lambda_U-2 \gamma
 		)^2-\gamma  \lambda_U \lambda_V^2\ln\mathfrak{a}}\right].
 	\end{equation}
 	Substituting \eqref{approx1B} in \eqref{approxH_EF} we obtain
 	\begin{equation}
 	\label{approxH_BD}
 	\mathcal{H}=\sqrt{\frac{U_0}{6\gamma}} \left(2\gamma-\lambda_U\right)^{\frac{\gamma -2 \lambda_U}{2\gamma}}\left(-2 c_1 (\lambda_U-2 \gamma
 	)^2-\gamma  \lambda_U \lambda_V^2\ln\mathfrak{a}\right)^{\frac{\lambda_U}{2\gamma}}.
 	\end{equation}
 	Integrating out for $\mathfrak{a}$ it follows
 	\begin{align}
 	&\mathfrak{a}=e^{\beta_2+\alpha_2  (\Gamma +\mathfrak{t})^{p_2}}\simeq e^{\alpha_2 \mathfrak{t}^{p_2}}\; (\text{leading terms as } \mathfrak{t} \rightarrow \infty),
 	\end{align}
 	where \\
 	$\alpha_2=-\frac{144^{\frac{\gamma }{\lambda_U-2 \gamma }} \left(\lambda_U \lambda_V^2 \left(-6^{\frac{\gamma }{2 \gamma
 				-\lambda_U}}\right) (2 \gamma -\lambda_U)^{1-\frac{\lambda_U}{\gamma }} \left(U_0 \left(2-\frac{\lambda_U}{\gamma }\right)\right)^{\frac{\gamma }{2 \gamma -\lambda_U}}\right)^{-\frac{2 \gamma }{\lambda_U-2 \gamma }}}{\gamma 
 		\lambda_U \lambda_V^2}$,\\ 
 	$\beta_2=\frac{c_1 \left(-\frac{8 \gamma }{\lambda_U}-\frac{2 \lambda_U}{\gamma }+8\right)}{\lambda_V^2},$ $\Gamma=\frac{\sqrt{6} c_2 (2 \gamma -\lambda_U)^{\frac{\lambda_U}{\gamma }}}{\sqrt{U_0 \left(2-\frac{\lambda_U}{\gamma
 			}\right)}},$ and $p_2=-\frac{2 \gamma }{\lambda_U-2 \gamma }.$ Notice that for $\lambda_U<0,\gamma>0$, we have $\alpha_2>0, 0<p_2<1.$ This implies that $\mathfrak{a}\rightarrow \infty$ as $\mathfrak{t}\rightarrow \infty$. 
 	
 	This implies that the Hubble parameter is
 	\begin{equation}
 	\mathcal{H}=\alpha_2 p_2 (\Gamma +\mathfrak{t})^{p_2-1}
 	\end{equation}
 	and the deceleration parameter is
 	\begin{equation}
 	\hat{q}=-1-\frac{(p_2-1) (\Gamma +\mathfrak{t})^{-p_2}}{\alpha_2 p_2}.
 	\end{equation}
 	Additionally, it is recovered the expected effective equation of state parameter, $\hat{w}_{\text{tot}}=-1$ as  $\mathfrak{t}\rightarrow \infty$ for $\gamma>0, \lambda_V\neq 0, \lambda_U<0$. 
 	
 	From the equations \eqref{FRIED2}, \eqref{approxH_EF} and \eqref{approx1B} we can obtain the following relation, which is valid at the critical point 
 	\begin{equation}
 	\label{POTENTIALBD}
 	V(\phi)=-\frac{U_0 \lambda_U}{2\gamma} e^{(2\gamma-\lambda_U)\chi}
 	=-\frac{ U_0 \lambda_U \left(-\frac{(\lambda_U-2 \gamma
 			)^2}{\gamma  \lambda_U \lambda_V^2 \ln (\mathfrak{a})+2 c_1
 			(\lambda_U-2 \gamma )^2}\right)^{2-\frac{\lambda_U}{\gamma
 			}}}{2 \gamma }.
 			\end{equation}
 			
   Therefore, using the equations \eqref{approx1B}, \eqref{approxH_BD}, and \eqref{POTENTIALBD}, we can obtain asymptotic expressions for $\chi, \mathcal{H}$ and $\phi$ in terms of $\mathfrak{t}$ after the substitution of $\mathfrak{a}\simeq e^{\alpha_2  \mathfrak{t}^{p_2}}$, which are valid as $\mathfrak{t}\rightarrow \infty$. 
	For this point \begin{align}
	&\mathfrak{R}=6\left(1+\frac{\mathfrak{a}\ddot{\mathfrak{a}}}{{\dot{\mathfrak{a}}}^2}\right)\left(\frac{\dot{\mathfrak{a}}}{\mathfrak{a}}\right)^2\nonumber\\
	&= 12 \alpha_2^2 p_2^2 (\Gamma +t)^{2 (p_2-1)}+6 \alpha_2 (p_2-1) p_2 (\Gamma +t)^{p_2-2},
	\end{align}
	and $\mathcal{H}=\alpha_2 p_2 (\Gamma +\mathfrak{t)}^{p_2-1},$
 where	$p_2=-\frac{2 \gamma }{\lambda_U-2 \gamma }.$ Notice that for $\lambda_U<0,\gamma>0$, we have $\alpha_2>0, 0<p_2<1.$ This implies that $\mathfrak{R}$ and $\mathcal{H}$ tends to zero  as $\mathfrak{t}\rightarrow \infty$ in such a way that $\mathfrak{R}/\mathcal{H}^2\rightarrow 12.$		
 			
 	Now, in order for $E_4$ to be a de Sitter solution it is required that $p_2=1$, which implies $\lambda_U=0$. But in this case the de Sitter solution will not be stable, rather, it will be a saddle, and the above approximation using the center manifold is not even valid. The other possibility for getting a de Sitter stage is to choose $\lambda_U<0$ and take the limit $\gamma\rightarrow +\infty$. In this case the solution would be stable, but this would imply $\omega_0\rightarrow -\frac{3}{2}$. This value of $\omega_0$ is several orders of magnitude lower than the bound $\omega_{0}>4\times 10^4$ imposed by the Solar System tests \cite{Will2014, BertottiIessTortora2003}, and the bounds estimated on the basis of cosmological arguments $\omega_{0}>120$ \cite{AcquavivaBaccigalupiLeachEtAl2005} and $10<\omega_{0}< 10^7$ \cite{NagataChibaSugiyama2004}.

\section{Concluding remarks}\label{SECT:5}

In this paper we have investigated the case of a modified JBD theory which includes a power-law potential for the JBD scalar field explicitly given by $U(\Phi)=U_0\Phi^{2-\frac{\lambda_U}{\gamma}}, \quad \gamma=(\omega_0+3/2)^{-1/2}$, where $\omega_0$ is the BD parameter, and a quintessence field with an exponential potential  $V (\phi) = V_0  e^{\lambda_V\phi}$, as the matter content. This scenario was analyzed in the Jordan and Einstein frames and equivalences between them were discussed.
We have presented the relation of our model
with the induced gravity model with power-law potential and it was discussed the integrability
of this kind of models when the additional quintessence field is massless, and has a small velocity. However, the main focus of this research was not to find analytical solutions as it was for the original induced gravity model, but
the study of the asymptotic behavior of the solution space without
using fine-tuning of the parameters and of the initial conditions. Nevertheless, the fact that for some fine-tuned
values of the parameters we may get some integrable cosmological models, makes our choice
of potentials very interesting. This issue deserves further investigation and it is left to future projects.

Secondly, using dynamical systems tools, we have provided the conditions on the parameters of the theory that lead to an attractor  with an effective equation of state parameter $w_{\text{tot}}=-1$,  describing the late-time evolution of the universe. The features of these solutions are essentially the same in both frames, i.e., analogous evolutions for the scale factor, and the conditions on the parameters are exactly the same. One condition is that the Brans-Dicke parameter satisfies $\omega_0>-\frac{3}{2}$, which is compatible with the ranges described by the observations, and the complementary condition is that $\lambda_U<0$, i.e., a power-law potential $U(\Phi)$ with a power greater than the second.  These conditions are independent  of the slope of the potential of the new scalar field, whenever its potential is not a constant.  We have proved that de Sitter solution is not the natural attractor in the JBD model. In this model, the attractor condition is $\gamma=\lambda_U$, which is forbidden in our scenario. Instead, we have shown that the attractor in the Jordan frame corresponds to an intermediate solution of the form 	$a(t)\simeq e^{\alpha_1  t^{p_1}}$ as $t\rightarrow \infty$ where $\alpha_1>0$ and $p_1=-\frac{2 \gamma }{\lambda_U-3 \gamma }$ with $0<p_1<1$ provided $\gamma>0, \lambda_U<0$. 
Furthermore, when we work in the Einstein frame we get that the attractor is also an intermediate solution of the form 	$\mathfrak{a}(\mathfrak{t})\simeq e^{\alpha_2  \mathfrak{t}^{p_2}}$ as $\mathfrak{t}\rightarrow \infty$ where $\alpha_2>0$ and $p_2=-\frac{2 \gamma }{\lambda_U-2 \gamma }$ with $0<p_2<1$ for $\gamma>0, \lambda_U<0$.
One possibility for getting a de Sitter stage is to choose $\lambda_U<0$ and take the limit $\gamma\rightarrow +\infty$. In this case the solution would be stable, but this would imply $\omega_{0}\rightarrow -\frac{3}{2}$, a value which is below of the lower limits on $\omega_0$ imposed by the Solar System tests, and the bounds estimated on the basis of cosmological arguments.

Furthermore, in the special case $\lambda_U=0$, that is, for a quadratic potential in the Jordan frame, or for a constant potential in Einstein's frame,  the above intermediate power-law solutions  are of a saddle type, which are not an attractor. These results were proved using the center manifold theorem, which is not based on linear approximation.  

Finally, for the specific elaboration of our extension of the induced gravity model in the Jordan frame, corresponding to the particular choice of a linear potential ${U}(\Phi)=U_0 \Phi$, $\lambda_U=\gamma=(\omega_0+3/2)^{-1/2}, 0<\gamma<\sqrt{\frac{2}{3}}$, the dynamical system is then reduced to a two dimensional one, and
the late-time attractor is linked with the exact solution found for the induced gravity model. This attractor solution satisfies the de Sitter-like asymptotic expansion $a \simeq e^{\frac{\left(2-\gamma
   ^2\right) t \sqrt{U_0}}{\sqrt{2}
   \sqrt{6-\gamma ^2}}}$ as $t\rightarrow \infty$, and the ``intermediate accelerated'' solution does not exist.
	
Summarizing, apart from these special fine-tuned examples (the linear, and quadratic potential ${U}(\Phi)$), it was shown that  ``intermediate accelerated'' solutions are generic late-time attractors in our
modified Jordan-Brans-Dicke theory.

\section*{Acknowledgments}

This work was funded by Comisi\'on Nacional de
Investigaci\'on Cient\'{\i}fica y Tecnol\'ogica (CONICYT) through: FONDECYT Grant 11110507
(A.C.), FONDECYT Grant 3140244 (G.L.), and by FONDECYT Grant 11140309 (Y.L.).
A.C. was partially supported by Universidad del Bio-Bio through grants DIUBB 121407 GI/VC, DIUBB 151307 and DIUBB GI150407/VC.


\providecommand{\href}[2]{#2}\begingroup\raggedright\endgroup

\end{document}